\documentclass{emulateapj}
\usepackage[english]{babel}     
\usepackage{amssymb,amsmath}    
\usepackage{graphicx}                   
\usepackage{braket}             
\usepackage{hyperref}
\usepackage{natbib}
\usepackage{bm}

\slugcomment{First submitted to ApJ: 19 August, 2015.}

\shorttitle{NGC~1277}
\shortauthors{Graham et al.}

\begin{document}

\title{A normal supermassive black hole in NGC~1277}

\author{Alister W.\ Graham, Mark Durr\'e, Giulia A.D.\ Savorgnan}
\affil{Centre for Astrophysics and Supercomputing, Swinburne University of
  Technology, Victoria 3122, Australia.}
\email{agraham@swin.edu.au}

\author{Anne M.\ Medling}
\affil{Research School of Astronomy and Astrophysics, Mount Stromlo Observatory,
Australia National University, Cotter Road, Weston, ACT 2611, Australia.}

\author{Dan Batcheldor}
\affil{Physics and Space Sciences Department, Florida Institute of Technology, 150
West University Boulevard, Melbourne, FL 32901, USA.}

\author{Nicholas Scott}
\affil{Sydney Institute for Astronomy (SIfA), School of
Physics, The University of Sydney, NSW 2006, Australia; 
Centre for Astrophysics and Supercomputing, Swinburne University 
of Technology, Victoria 3122, Australia.}

\author{Beverly Watson} 
\affil{Physics and Space Sciences Department, Florida Institute of Technology, 150
West University Boulevard, Melbourne, FL 32901, USA.}

\and

\author{Alessandro Marconi} 
\affil{Dipartimento di Fisica e Astronomia, Universit\`a di Firenze, via
  G. Sansone 1, I-50019 Sesto Fiorentino, Firenze, Italy.}

\begin{abstract}

The identification of galaxies with `overly massive' black holes 
requires a black hole mass ($M_{\rm bh}$) {\it and} a host
spheroid mass ($M_{\rm sph,*}$). Here we provide our measurements for NGC~1277. 
Our structural 
decomposition reveals that NGC~1277 is dominated by a `classical'
spheroid with a S\'ersic index $n=5.3$, a half-light radius $R_{\rm e,major}=2.1$ kpc, and a stellar
mass of $2.7\times10^{11}~M_{\odot}$ (using $M_*/L_V=11.65$, 
Mart\'in-Navarro et~al.).  This mass is an order of magnitude greater than originally reported. 
Using the latest $M_{\rm{bh}}$--$n$,
$M_{\rm{bh}}$--$M_{\rm{sph,*}}$, and 
$M_{\rm{bh}}$--$\sigma$ 
relations, the expected black hole mass is respectively
($0.57^{+1.29}_{-0.40})\times10^9~M_{\odot}$, 
($1.58^{+4.04}_{-1.13})\times10^9~M_{\odot}$, and 
($2.27^{+4.04}_{-1.44})\times10^9~M_{\odot}$ (using $\sigma=300$ km s$^{-1}$) for
which the `sphere-of-influence' is $0\arcsec.31$.  Our new kinematical 
maps obtained from laser guide star assisted, adaptive optics on the Keck I
Telescope dramatically reaffirm the presence of the inner, nearly edge-on, disk 
seen in this galaxy's image.  We also report that this 
produces a large velocity shear ($\sim$400~km~s$^{-1}$) across the inner $0\arcsec.2$
(70~parsec) plus elevated values of $\sqrt{\sigma^2+V_{rot}^2}$ across the 
inner $(\pm1\arcsec.9)\times(\pm0\arcsec.3)$ region of the galaxy. 
Our new multi-Gaussian expansion (MGE) models 
and Jeans Anisotropic MGE (JAM) analysis struggled to match this component. 
Our optimal black hole mass, albeit a probable upper limit because of the disk, is 
$1.2\times10^{9}~M_{\odot}$ ($M/L_V=12.3$). This is 
an order of magnitude smaller than originally reported and 4 times smaller 
than recently reported.  It gives 
an $M_{\rm{bh}}/M_{\rm{sph,*}}$ ratio of 0.45\% in agreement
with the median ($\approx$0.5\%) and range (0.1--5.0\%) observed in non-dwarf,
early-type galaxies. This result 
highlights the need for caution with inner disks. 




\end{abstract}


\section{Introduction}

Using the Hobby-Eberly Telescope's Marcario spectrograph under 1$\arcsec$.6
seeing conditions, van den Bosch et al.\ (2012) claimed to have discovered a
$1.7\pm0.3\times10^{10}~M_{\odot}$ black hole in the nearby, compact but
massive early-type galaxy NGC~1277.  Not only is this the largest reported  black hole
mass with a direct measurement, but they remarked that it weighs in at an
extraordinary 59\% of its host spheroid's stellar mass, and 14\% of the galaxy's
total stellar mass.  To put this in perspective, Graham \& Scott (2013)
reported that the average $M_{\rm bh}/M_{\rm sph,*}$ mass ratio was
0.49\% for large spheroids, a result reiterated by Kormendy \& Ho (2013), 
and recently upgraded to 0.68\% by Savorgnan et al.\ (2015) using 
multi-component decompositions for a sample of 66 galaxies imaged in the
infrared. 

Due to this surprising black hole mass in NGC~1277, Emsellem (2013)
revisited its derivation using the same data.  
Preferring an {\it F550M} (narrow $V$-band) stellar mass-to-light 
ratio of 10 $M_{\odot}/L_{\odot}$ 
(consistent with a Salpeter-like Initial Mass Function) 
rather than the value of $\sim$6 $M_{\odot}/L_{\odot}$ reported in van den
Bosch et al.\ (2012), Emsellem suggested a lower black hole mass of 
$0.5\times10^{10}~M_{\odot}$, albeit with a best fitting range reaching up to 
2.5$\times10^{10}~M_{\odot}$ due to his advocation for expanding the
confidence intervals. 
Y{\i}ld{\i}r{\i}m et al.\ (2015) subsequently reported a preferred
black hole mass of $1.26^{+0.32}_{-0.47}\times10^{10}~M_{\odot}$
($M_*/L_{V,sph}=6.5\pm1.5~M_{\odot}/L_{\odot}$), while Scharw\"achter et
al.\ (2015) reported a value consistent with $1.7\times10^{10}~M_{\odot}$
($M/L_{V,sph}=6.3~M_{\odot}/L_{\odot}$) based on the kinematics of CO(1-0)
emission measured with the IRAM Plateau de Bure Interferometer at
$\sim$1-arcsec resolution.  
Most recently, Walsh et al.\ (2015) have confirmed the optimal mass derived 
by Emsellem (2013), reporting $M_{\rm bh} = 4.9\pm1.6\times10^9~M_{\odot}$
and $M_*/L_{V,sph} = 9.3\pm1.6~M_{\odot}/L_{\odot}$. 
All of these measures still leave a galaxy with an 
unusually high (black hole)-to-spheroid mass ratio.   
Emsellem (2013) additionally questioned whether the spheroid had been properly
identified by van den Bosch et al.\ (2012), and he suggested that there may be
a spheroid with a mass of $\sim$$1.8\times10^{11}~M_{\odot}$, or
$\sim$$10^{11}~M_{\odot}$ within the inner 2 kpc, obtained by summing all of
the components of his Multi-Gaussian photometric Expansion (MGE) model which
had an ellipticity less than 0.7.  Coupled with his reduced black hole mass,
it gave an $M_{\rm bh}/M_{\rm sph,*}$ mass ratio of 3--5\%.

We start our investigation of NGC~1277 by presenting an 
image analysis and measurement of the physical properties 
of its spheroidal stellar system.  
We 
perform a decomposition of NGC~1277 based on fundamental galaxy components. 
Knowing the properties of the spheroid, such as its S\'ersic index and stellar
mass, not only enables us to determine the {\it expected} black hole 
mass, but provides some insight into 
the history of NGC~1277.  As suggested in Graham (2013) 
and Driver et al.\ (2013), the bulges in some of today's 
massive lenticular galaxies are likely to be the descendents of the compact galaxies
seen at high-redshift\footnote{Secular evolution of disks does not build
  massive bulges with S\'ersic indices $n>2$, and minor mergers will evolve, rather
  than preserve, the compact high-$z$ galaxies.}. 
Many have the same small sizes, large masses, and 
radial concentrations of light --- as traced by the S\'ersic index (Dullo \&
Graham 2013).  The subsequent accretion and growth of disks (e.g.\ White \&
Rees 1978; White \& Frenk 1991), perhaps from cold gas
flows (e.g.\ Pichon et al.\ 2011; Combes et al.\ 2014), around the initially
compact galaxies would increase their total galaxy size, effectively hiding
them --- today's massive bulges --- in plain sight (Graham, Dullo \& Savorgnan
2015).  
%
In NGC~1277, as in other nearby galaxies like NGC~1332 ($R_{\rm e} = 2.0$ kpc, 
$M=1.1\times 10^{11}~M_{\odot}$, Savorgnan \& Graham 2015b), 
the disk is small enough that the whole galaxy is still a compact massive
galaxy (see section~\ref{Sec1D}).

Accounting for the different structural components in a galaxy can also be
important for the direct measurement of the black hole mass, in particular
with regard to the influence of disks.  Unresolved disk rotation can enhance
the observed velocity dispersion within a given resolution element, an issue
noted by Tonry (1984) in regard to M32 and by Dressler (1989, his section~3)
in regard to M87.  Such ``velocity shear'' is also known to be an important
consideration when dealing with poorly resolved galaxies at large redshift,
where rotation can artificially raise the observed velocity dispersion
(e.g.\ Law et al.\ 2009; Green et al.\ 2014).  As noted in the conclusions of
Graham et al.\ (2011), exactly the same situation can occur at the centers of
nearby galaxies if they possess a (nearly) edge-on disk. 
Given that NGC~1277 has both an intermediate-scale disk (see Section~2) and a
highly-inclined inner disk (Sections~2 and 5), this is of particular concern.  
%
%
Using new, high spatial-resolution
spectroscopic data from the {\sc osiris} instrument on the Keck I Telescope,
in the second part of this paper we explore the spataila extent and 
kinematics of this inner disk. We additionally attempt an independent
measurement of the galaxy's black hole mass. 
%

%

Our paper is laid out as follows.  In Section~\ref{Sec_Image} we perform a
careful structural decomposition of a {\it Hubble Space Telescope} image of
NGC~1277.  Using the results from this analysis, in Section~\ref{Sec_predict}
we derive the expected black hole mass in NGC~1277.  In Section~\ref{Sec_Kin}
we present our new kinematic data from {\sc osiris}, and in
Section~\ref{Sec_Mass} we present a derivation of the black hole mass.
Finally, we provide a discussion, including relevance to galaxies other than
NGC~1277, and present our conclusions in Section~\ref{Sec_DC}. 

Using the Planck 2013 results, we assume a spatially flat Universe
(i.e.\ $\Omega_m + \Omega_{\Lambda} =1$) with 
a cosmological matter fraction $\Omega_m = 0.315\pm0.017$ 
and a Hubble constant $H_0 = 67.3\pm1.2$ km s$^{-1}$ Mpc$^{-1}$ 
(Planck Collaboration 2014).  Correcting the
heliocentric radial velocity of 5066$\pm$28 km s$^{-1}$ (Falco et al.\ 1999)
for (Virgo + Great-Attractor + Shapley) infall gives a redshift of 4983$\pm$33
km s$^{-1}$ (obtained courtesy of
NED\footnote{\url{http://nedwww.ipac.caltech.edu}}) and an `angular size distance' of
72.5 Mpc when using the above cosmology. 
This corresponds to a scale of 352 parsec per arcsecond (Wright 
2006)\footnote{\url{http://www.astro.ucla.edu/~wright/CosmoCalc.html}}.
The `luminosity distance' is 75.0 Mpc, giving a cosmological redshift corrected
distance modulus of 34.38 mag.
For reference, this is practically identical to the value of 34.39 mag (and
$1\arcsec = 353$ pc) used by van den Bosch et al.\ (2012).

\section{Image Analysis}\label{Sec_Image}

We use an archival {\it Hubble Space Telescope (HST)}, {\it Advanced Camera
  for Surveys (ACS}, Ford et al.\ 1998)
observation\footnote{\url{http://archive.stsci.edu/hst/search.php}} taken through an
{\it F550M} (narrow $V$-band) filter as a part of Proposal Id.\ 10546 (Canning
et al.\ 2010).  In data set J9BB01040, NGC~1277 is located off to the side of
those authors' primary target galaxy NGC~1275, located some 4 arcminutes
away. 

\begin{figure*}
\begin{center}
\includegraphics[width=0.4\textwidth]{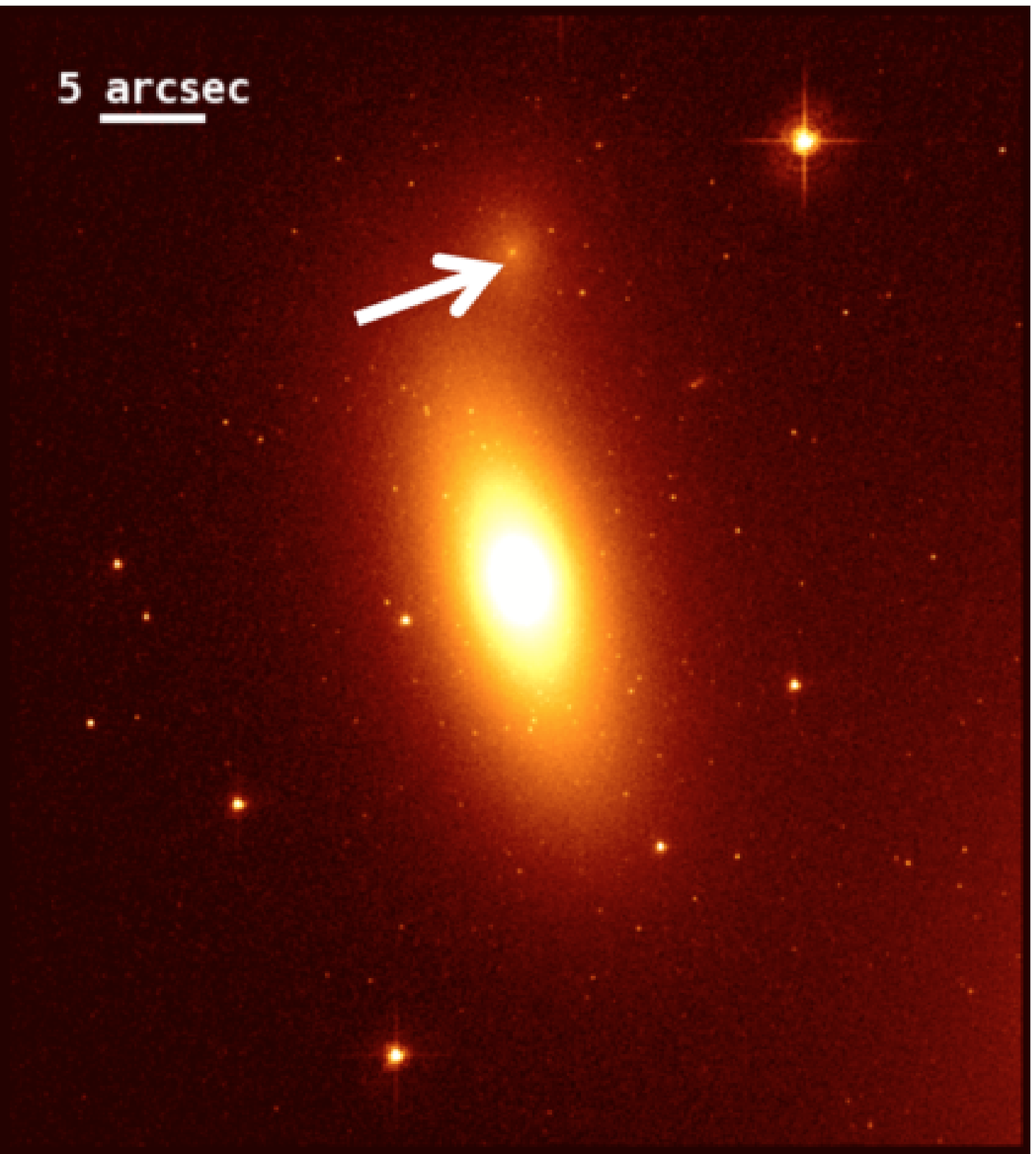} 
\includegraphics[width=0.4\textwidth]{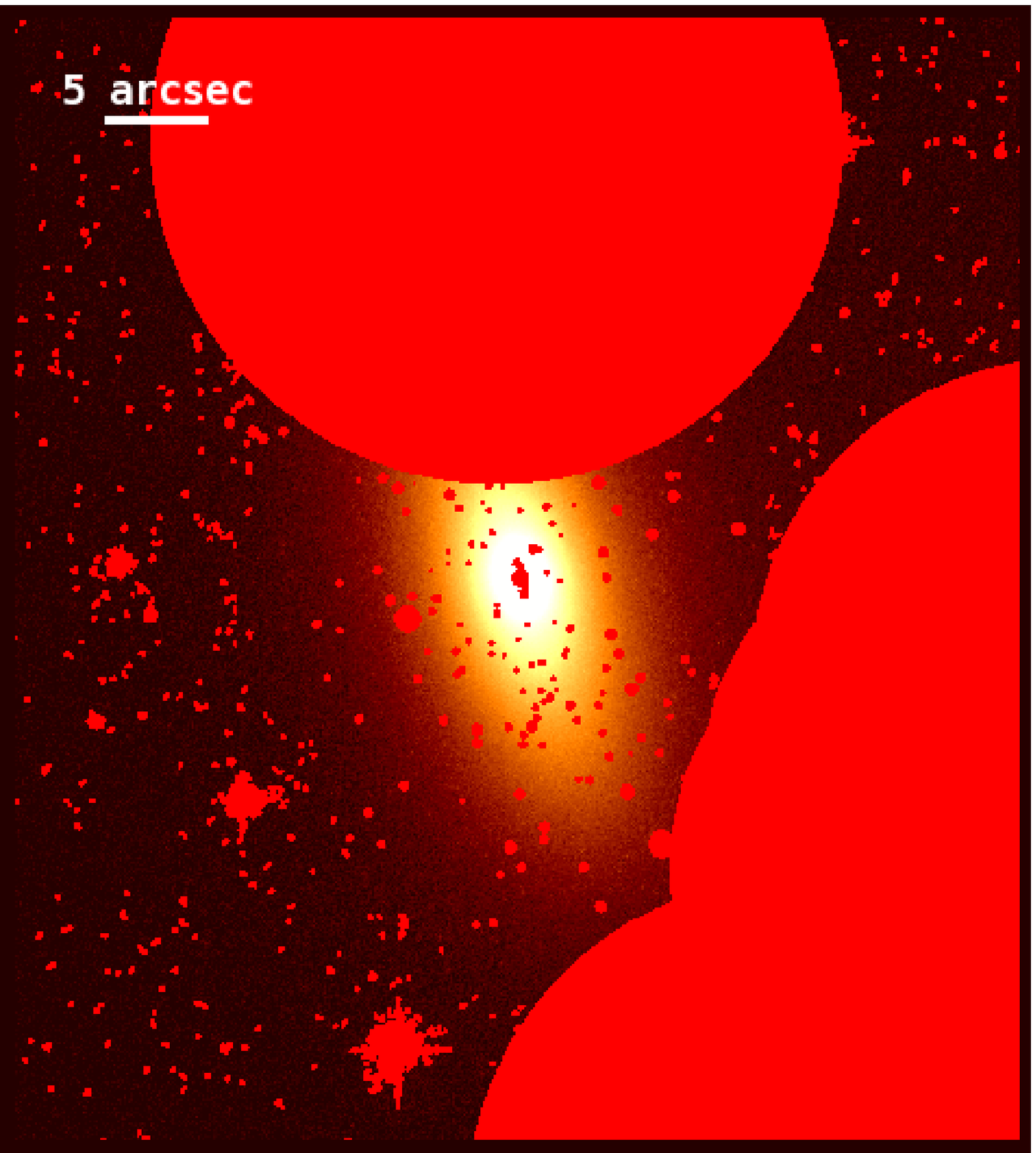} 
\caption{Left: Cut-out postage-stamp {\it HST F550M} image showing NGC~1277. 
  The arrow points in a north-to-south direction (east is roughly down) 
  and marks the location of the galaxy that was masked
  as a part of our image analysis.  The glow of NGC~1278 can be seen in the
  lower-right. 
  The sky-background was measured outside of this cut-out area. 
  Right: {\it HST F550M} image of NGC~1277 with mask.} 
\label{fig:image}
\end{center}
\end{figure*}

The background sky flux was subtracted to give a count of zero in the
galaxy-free corners of the image.  Galactic stars and background galaxies in
the frame have been masked, and thus avoided in our analysis.  We
also paid special attention to the masking of two neighboring galaxies.  The
outskirts of NGC~1278 --- a peculiar elliptical galaxy --- appears as faint halo
light in the bottom-right corner of Figure \ref{fig:image}.  The second 
neighbor is an elliptical galaxy (indicated with a white arrow in Figure
\ref{fig:image}) sitting above NGC~1277 at a projected separation of
$\sim 15\arcsec$.  These two galaxies have been heavily masked by us, ensuring that
no significant light was affecting the outer regions of NGC~1277.  Finally,
our masking includes the near side of an obvious dust lane/ring affecting the nuclear
region of NGC~1277.  The final mask is shown in Figure \ref{fig:image}.

\subsection{1D modelling}\label{Sec1D}

\begin{figure}[!h]
\includegraphics[trim=0cm 1.4cm 0cm 1.8cm, width=0.537\textwidth]{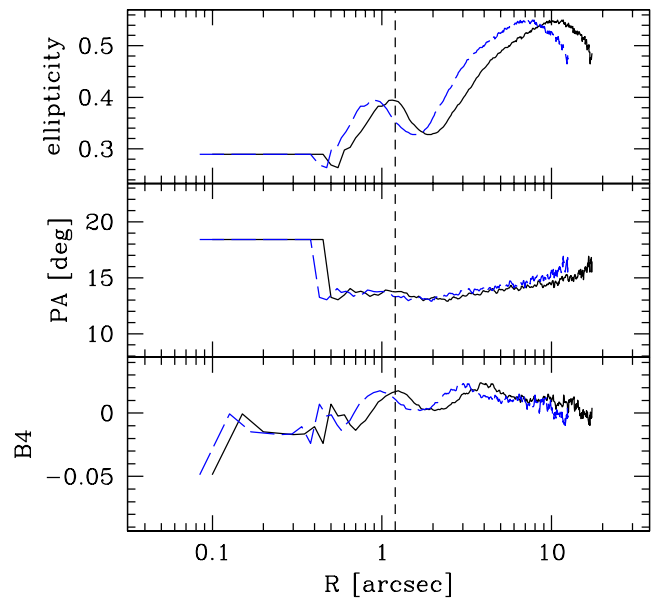} 
\caption{Ellipticity (\emph{top panel}), position angle 
(\emph{middle panel}) and fourth harmonic (\emph{bottom panel}) 
radial profiles extracted along the major-axis (black solid line) 
and the equivalent-axis (blue dashed line).  The constant (horizontal) value inside
of $\sim$0$\arcsec$.5 was set by the IRAF task {\tt ellipse} and is not real. 
The vertical black dashed line at $R=1\arcsec$.2 
indicates the position of a peak in the ellipticity and fourth harmonic 
profiles (along the major-axis), revealing the presence of an embedded
component.  The second peak in the ellipticity profile at $\sim$10$\arcsec$
reveals the radius where the intermediate-scale disk contributes most light 
(relative to the rest of the galaxy). 
} 
\label{fig:isoph}
\end{figure}

After masking the galaxy image, we performed an isophotal analysis with the
\emph{IRAF} task {\tt ellipse} (Jedrzejewski 1987)\footnote{Our analysis was
  performed before {\tt isofit} (Ciambur 2015) was conceived or available.}, 
holding the centre fixed and allowing the
position angle and ellipticity to vary with radius.
Figure~\ref{fig:isoph} presents the ellipticity, `position angle' and `fourth
harmonic'\footnote{Quantifying the deviations of the isophotes from perfect
  ellipses using a Fourier series (Carter 1978), we show the amplitude of the
  coefficient of the $\cos(4\theta$) term.} 
radial profiles along the major-axis, and also along the
equivalent-axis, i.e.\ the geometric mean of the major ($a$) and minor ($b$)
axis ($R_{\rm eq} = \sqrt{ab}$) equivalent to a kind of circularized profile.
The peaks at $R \sim 1\arcsec$.2 in the ellipticity and fourth harmonic
major-axis profiles are an immediately evident feature, indicating the
presence of an inner disky component that has largely declined by $2\arcsec$.
10-12$\arcsec$ along the major-axis, and at 7-8$\arcsec$ along the
equivalent-axis, clearly signaling the presence of an intermediate-scale
disk. 
There are therefore two additional components apparent in NGC~1277, i.e.\ in
addition to the main spheroidal structure.

In passing we note that galaxies with intermediate-scale disks, i.e.\ galaxies
which are intermediate between `elliptical' (E) and `lenticular' (S0)
galaxies, are not unusual.  They were first referred to as ES galaxies by Liller
(1966) and as E/S0 galaxies by Strom et al.\ (1977; see also Strom \& Strom 1978
and Thompson 1976).  They have since been referred to as ``S0-like'' (Michard
1984) and ``disk-ellipticals'' (Nieto et al.\ 1988) or ``disky ellipticals''
(Simien \& Michard 1990).  Unlike S0 galaxies, ES galaxies do not have
extended disks which dominate the light at large radii.

\begin{figure*}
\includegraphics[trim=0cm 2cm 0cm 1.5cm, width=1.\textwidth]{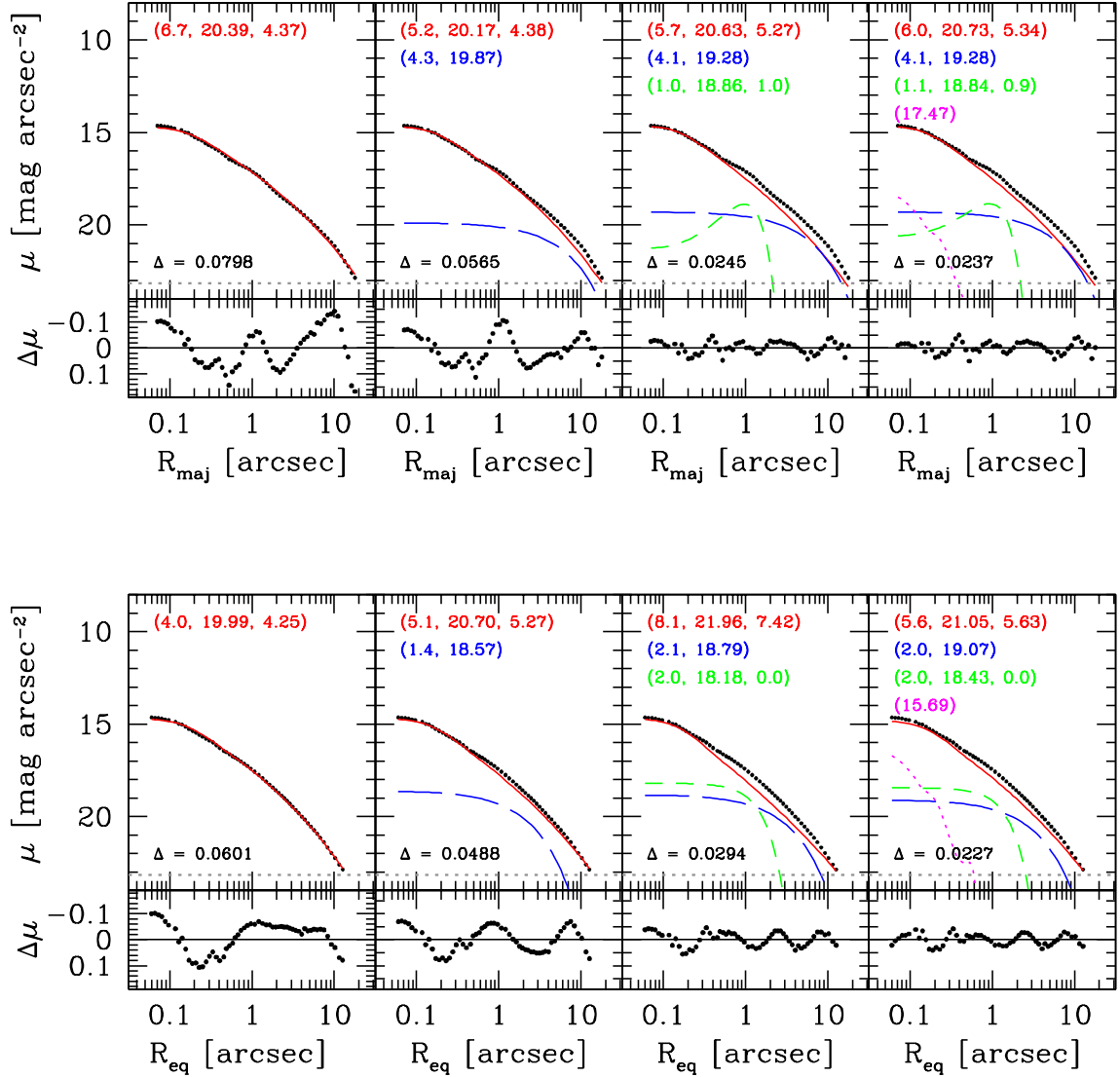} 
\caption{
{\it F550M} surface brightness profile (in units of $\rm mag~arcsec^{-2}$) along the
major-axis (\emph{top panels}) and the equivalent-axis (\emph{bottom
  panels}).  From left to right, the panels of
each column show a fit to the surface brightness profile with an increasing
number of (PSF-convolved) model components: a S\'ersic profile (red solid line), an
exponential disk (blue long-dashed line), a Gaussian ring (green short-dashed
line) and a PSF profile (pink dotted line).  
The best-fit parameters are displayed with the same color
coding and are as follows: 
$(R_{\rm e},~\mu_{\rm e},~n)$ for the S\'ersic model,
$(h,~\mu_{0})$ for the exponential model,
(FWHM,$~\mu_{0},~R_{0}$) for the Gaussian ring model
and $(\mu_{0})$ for the PSF model.
As discussed in the text, the number of model components was additionally guided by
the ellipticity profile and the deviation of the isophotes from ellipses at
different radii (Figure~\ref{fig:isoph}). 
The horizontal grey dashed line indicates 3 times the $RMS_{\rm sky}$ level,
and $\Delta = \sqrt{ \sum_{i=1,N} ({\rm data}_i-{\rm model}_i)^2/(N-\nu)}$, 
where $N$ is the number of data points and $\nu$ is the number of model 
parameters involved in each fit. 
} 
\label{fig:fit}
\end{figure*}

We fit the major- and equivalent-axis surface brightness profiles with a
PSF-convolved model, where the radial point spread function (PSF) has been
measured from suitably bright stars in the image and described with a Moffat
function ($FWHM = 0\arcsec .13$, $\beta=2.9$).  In Figure~\ref{fig:fit} we
report a slight variation of the frequently used ``root-mean-square
deviation'' ($= \sqrt{ \sum_{i=1,N} ({\rm data}_i-{\rm model}_i)^2/N}$) when
quantifying the global goodness of each fit.  Our variation is that we used
$N-\nu$ rather than simply $N$ in the denominator due to our successive
fitting of models with increased numbers of free parameters $\nu$.  This does
not change the actual fit in any way, and with $N=58$ and $\nu$ ranging from
just 3 to 9, the differences are small.  For example, the fits to the
equivalent-axis (lower panels in Figure~\ref{fig:fit}) have ``root-mean-square
deviation'' values of 0.0585, 0.0466, 0.0273 and 0.0209.  It should also be
kept in mind that both of these measurements reflect the {\em global} goodness
of fit, and not
the local goodness of fit.  
For example, adding a point-source will not have a big impact 
on the global ``root-mean-square deviation'' even though it may considerably
reduce the residuals on a local scale where this component has been added. 
We have not used the (reduced) chi-squared statistic because this quantity is
only useful when one has meaningful measurement errors.  Due to dust, PSF 
errors, possible additional components, one may place an inappropriate weight
on the data near the centre when using a (signal-to-noise)-weighted fitting
scheme that does not readily account for these biasing factors. 

As already noted, NGC~1277 is
not a pure elliptical galaxy.  Indeed, elliptical galaxies with ellipticity
greater than 0.5, i.e.\ the old E5, E6 and E7 elliptical galaxies, were
recognized as lenticular disk galaxies many decades ago (e.g., Liller 1966 and
Vorontsov-Vel'Yaminov \& Arkhipova's 1962--1968 ``Morphological catalogue of
galaxies'').  Consequently, a single S\'ersic model (shown by the left-most
panels in Figure~\ref{fig:fit}) does not provide an optimal description of
NGC~1277's light profile, and this is evident by the high amplitude of the
residual profile.  The addition of an exponential disk component (second
column) improves the fit, but the residuals are still affected by the
previously identified structure peaking at $R \sim 1\arcsec .2$ along the
major-axis (and at $R \sim 1\arcsec .0$ along the equivalent-axis) of the
ellipticity profile (Figure~\ref{fig:isoph}).  While this is likely to be an
inner disk --- a feature that is common in galaxies (e.g.\ Rest et al.\ 2001;
B\"oker et al.\ 2002; Balcells et al.\ 2007; Seth et al.\ 2008; Graham et
al.\ 2012) --- when fitting the major-axis light profile, we found that the
best description of this component is obtained with a Gaussian ring model
(third column).  This is perhaps not unexpected given the obvious dust ring
(with a major-axis near 0.7$\arcsec$, and a width of $0.\arcsec
2$)\footnote{The dust ring peaks along the minor axis at $0.\arcsec 2$.}.
Adding a central point source, represented by the {\it HST} PSF and shown in
the fourth column of Figure~\ref{fig:fit}), provides a better fit although we
note that this is probably not an AGN (Fabian et al.\ 2013).  Along the
major-axis, the Gaussian ring model peaks at $0.\arcsec 9$ and has a FWHM
width of $1.\arcsec 1$.

This combination of a point-source and a ring, rather than a single
exponential model for an inner disk, is likely due to the ring of dust
embedded in the suspected inner disk.  The mask that we used to exclude the
dusty region was already at a maximum; expanding it further resulted in an
insufficient amount of image for the task {\tt ellipse}, and thus it crashed
and failed to produce a light-profile when the mask was extended.  Along the
equivalent-axis, the peak of the dust ring is closer to the center, and the
peak of our Gaussian ring model when fit to the equivalent-axis 
light-profile is at $R=0$; that is, the Gaussian ring has reduced to a normal
Gaussian, which in this instance has a FWHM equal to $2\arcsec$ (see the lower
right panels of Figure~\ref{fig:fit}).  Together, these two components (a point
source plus a PSF-convolved, extended Gaussian centered at $R=0$) combine to
approximate the inner disk.  It turns out that 
this disk is clearly revealed in the kinematic map shown in Section~4, in
addition to the ellipticity profile shown in Figure~\ref{fig:isoph}. 
Table~1 provides a summary of the 
best-fitting parameters for each component fit to the galaxy. 
Again, we did not use a signal-to-noise weighted fitting scheme which 
would have  
heightened the sensitivity of the fit to the problematic dusty inner region and 
amplified any PSF mismatch, thereby driving the fit away from the true 
solution. To mitigate against these concerns, an equal weighting was placed on
the data shown in Figure~\ref{fig:fit}, as is commonly done when modelling
galaxy surface brightness profiles. 

In place of the ring model, we tested the use of an inner exponential
component, and alternatively an inner S\'ersic component (because nearly
edge-on disks are better described by a S\'ersic model with index less than 1,
e.g.\ Pastrav et al.\ 2013a,b), but the solution was unsatisfactory and the
data favoured the ring model. 
Integrating the flux of the ring model fit
to the equivalent-axis light profile data (which is simply a Gaussian 
in this instance) gives an observed apparent 
magnitude of 16.76 mag (AB mag, {\it F550M}).  Accounting for $A_{\rm
  V} = 0.452\rm~mag$ of Galactic extinction (Schlafly \& Finkbeiner
2011, via the NASA/IPAC Extragalactic Database
(NED)\footnote{\url{http://nedwww.ipac.caltech.edu}}), gives a
corrected apparent magnitude of 16.31 mag (AB system).  We have not
attempted any evolutionary nor $K$-correction.  At a `luminosity
distance' of 75 Mpc, and using $M_*/L_V = 11.65$ (see
section~\ref{Sec_Mass} and Mart\'in-Navarro et al.\ 2015a), this
equates to a stellar mass of 1.67$\times10^{10}~M_{\odot}$.  
Given the 
lack of a significant AGN in this galaxy (Fabian et al.\ 2013), and
given that the central point source flux equates to a stellar mass of
$\sim \times10^9~M_{\odot}$ (see Table~1), the central point source is too large to be a
spheroidal star cluster (e.g.\ Scott \& Graham 2013) and is therefore
most likely part of the inner disk.  Adding this flux to the 
`Gaussian ring' gives a combined mass of 1.8$\times10^{10}~M_{\odot}$.

This is a massive inner disk (an order of magnitude greater than the
smaller-scale nuclear disks reported in Balcells et al.\ 2007 and Scott \&
Graham 2013).  It may therefore be expected to have a large rotation.  An
interesting, alternative, hypothesis is that this feature may instead be a bar
seen close to end-on (Emsellem 2013).  If this was the case, then the
inclination of the intermediate-scale disk (within which this bar must reside)
as determined from the geometry of the dust (presumably now in the bar), which
is not circular when seen face-on, could be too low.  Although, one might
expect a disk which is closer to edge-on in NGC~1277 to be more apparent at
intermediate radii because of the $2.5\log(b/a)$ brightening of its surface
brightness (in the absence of dust) as the apparent axis-ratio $b/a$ goes from
1 to 0 (e.g., Graham 2001).  In the following subsection we perform a 2D image
analysis to better measure the axis-ratio of the intermediate-scale disk and
thereby check for a disk more inclined than $\sim$74 degrees.

The total {\it F550M} galaxy magnitude is $13.49 \rm~mag$.  Correcting for
Galactic dust, this brightens to 13.04 mag, and the (Galactic dust)-corrected
bulge magnitude is 13.29 mag (AB, {\it F550M}).  NGC~1277 has a classical
bulge with a S\'ersic index $n_{\rm maj} = 5.34$ that accounts for 79.4\% of
the total galaxy light.  The intermediate-scale disk's contribution is 15.2\%,
while the inner ring / inner disk component makes up $\sim5$\%.  The
percentage by mass is slightly different based on the stellar mass-to-light
ratios used for each component (Table~1).

\begin{table*}
\label{Tab_Comp}
\centering 
\caption{Component parameters.}
\begin{tabular}{lcccccl}
\hline\hline 
Component  &  size            &  $\mu$             & S\'ersic        &  mag   & $M_*/L_V$             & Mass \\
           & [$\arcsec$ / kpc] & [mag arcsec$^{-2}$] & index           &  [mag] & [$M_{\odot}/L_{\odot}$] & [$M_{\odot}$] \\  
\hline
NGC~1277   &    ...          & ...                & ...             & 13.04  &  ...             & 3.00$\times10^{11}$ \\
\hline 
Spheroid   & $R_{\rm e}$       &  $\mu_{\rm e}$       &                 &       &                   &                    \\
           & 6.0 / 2.12      &  20.73              &  5.34           & 13.29 &  11.65            & 2.69$\times10^{11}$ \\ 
\hline
Int.-scale disk  &   $h$           & $\mu_0$             &                 &       &                   &    \\
           & 4.1 / 1.45      &  19.28              &  1.0            & 15.09 &  3.00             & 1.32$\times10^{10}$ \\
\hline 
Nucleus    & $FWHM$          &  $\mu_0$            &  ...            &       &                   &    \\
           & 0.13 / 0.046    &  17.47              &   ...           & 19.21 & 11.65             &    1.15$\times10^9$ \\
\hline\hline
           & $R_{\rm peak}$    & $\mu_{\rm peak}$       &   width         &  Mag  & $M_*/L_V$              &  Mass    \\
           & $\arcsec$ / kpc & [mag arcsec$^{-2}$]  & $\arcsec$ / kpc & [mag] & [$M_{\odot}/L_{\odot}$] &  [$M_{\odot}$]  \\  
\hline
Inner ring & 0.9 / 0.32      & 18.84               & 1.1 (0.39)      & 16.31 &  11.65             &  1.67$\times10^{10}$ \\
\hline 
\end{tabular}

Best-fitting (major-axis) structural parameters from the upper right panel of
Fig.~\ref{fig:fit}.  The spheroidal component has been fit with a S\'ersic
model, the intermediate-scale disk with an exponential model of scale-length
$h$, and the inner disk with the combination of an `inner ring' (a Gaussian
function centred at $R_{\rm peak}=0\arcsec .9$) plus the {\it HST} point
spread function for its nucleus.  The observed dust ring effectively broke the
inner disk into two components.  The (uncorrected) surface brightnesses are in
units of {\it F550M} mag arcsec$^{-2}$.  The apparent magnitudes (which have
been corrected for Galactic extinction) also relate to the same narrow
$V$-band {\it F550M} filter.  Note that these magnitudes were calculated using
the best-fitting model components to the equivalent-axis light profile (lower
right panel of Fig.~\ref{fig:fit}) and integrating to $R=\infty$ while
assuming circular symmetry.  The stellar mass-to-light ratios are explained in
subsection~\ref{Sec_Mass}.  The masses are in units of solar masses and were
obtained using a distance modulus of 34.38 mag and $M_{{\rm V},\odot}=4.82$
mag (Cox 2000).  
\end{table*}

Given that galaxy disks typically have fixed ellipticity, reflecting their
inclination to our line of sight, the peak in the ellipticity profile at $R
\sim 10\arcsec$ along the major-axis (Figure~\ref{fig:isoph}) can be
interpreted in the following way.  Going from the galaxy centre to the
outskirts, the disk light becomes increasingly important relative to the spheroid's
light, reaching its maximum at $R \sim 10\arcsec$.  Beyond $R \sim 10\arcsec$,
the contribution from the disk light starts declining more rapidly than the
spheroid light (e.g.\ Liller 1966). 
This suggests a somewhat embedded, intermediate-size stellar
disk for NGC~1277 (see Savorgnan \& Graham 2015c for other examples).  
This explanation is also in accord with the results from the model fitting
seen in Figure~\ref{fig:fit}.  In passing we note that after a careful test,
we ruled out the possibility that the decline in the ellipticity profile
beyond $10\arcsec$ is caused by contaminating light from the (masked)
neighboring galaxies.  The peak ellipticity 
($\epsilon = 1-b/a = 1-\cos i$) at 10$\arcsec$ of 
0.54 gives a minimum inclination, from face-on, of 63 degrees. This is a
lower limit because this ellipticity is still somewhat diluted by the flux
from the spheroidal component of this galaxy.  Following van den Bosch et
al.\ (2012), they suggested that the inner dust disk, with its observed axis
ratio of $\sim$0.3, may in fact be circular and reside in the plane of the
larger stellar disk, which they subsequently took to have an inclination of 75
degrees.  Based on the ellipticity of the dust ring, and assuming that it
would appear circular when seen face-on, we too derive an inclination 
of 74$\pm$2 degrees for the inner disk within which the dust ring is embedded.

\subsection{2D modelling}\label{Sec2D}

Fitting a two-dimensional (2D) model to the image was performed with the code
{\tt Imfit} (Erwin
2015)\footnote{\url{http://www.mpe.mpg.de/~erwin/code/imfit/}}, the
previously created mask, and a Tiny Tim PSF (Krist 1995). 
The optimal model was found to consist of a 
S\'ersic-spheroid, an intermediate-scale exponential disk, plus a
centrally-located Gaussian (equivalent to a S\'ersic model with an index $n=0.5$) 
for the suspected inner disk.  This is consistent with our
fit to the equivalent-axis light profile discussed earlier.  
The innermost 
data point was masked in our 2D modelling and no central point-source was required. 
As before, we were not able to obtain a satisfying fit using an exponential
model for the inner disk (identified as such from the kinematics, see
later), however it is common for (nearly) edge-on disks to be well described by a 
S\'ersic model with an index less than 1 (e.g., Pastrav et al.\ 2013a,b).

\begin{figure}[!h]
\includegraphics[trim=0cm 0cm 0cm -1.3cm, width=0.23\textwidth]{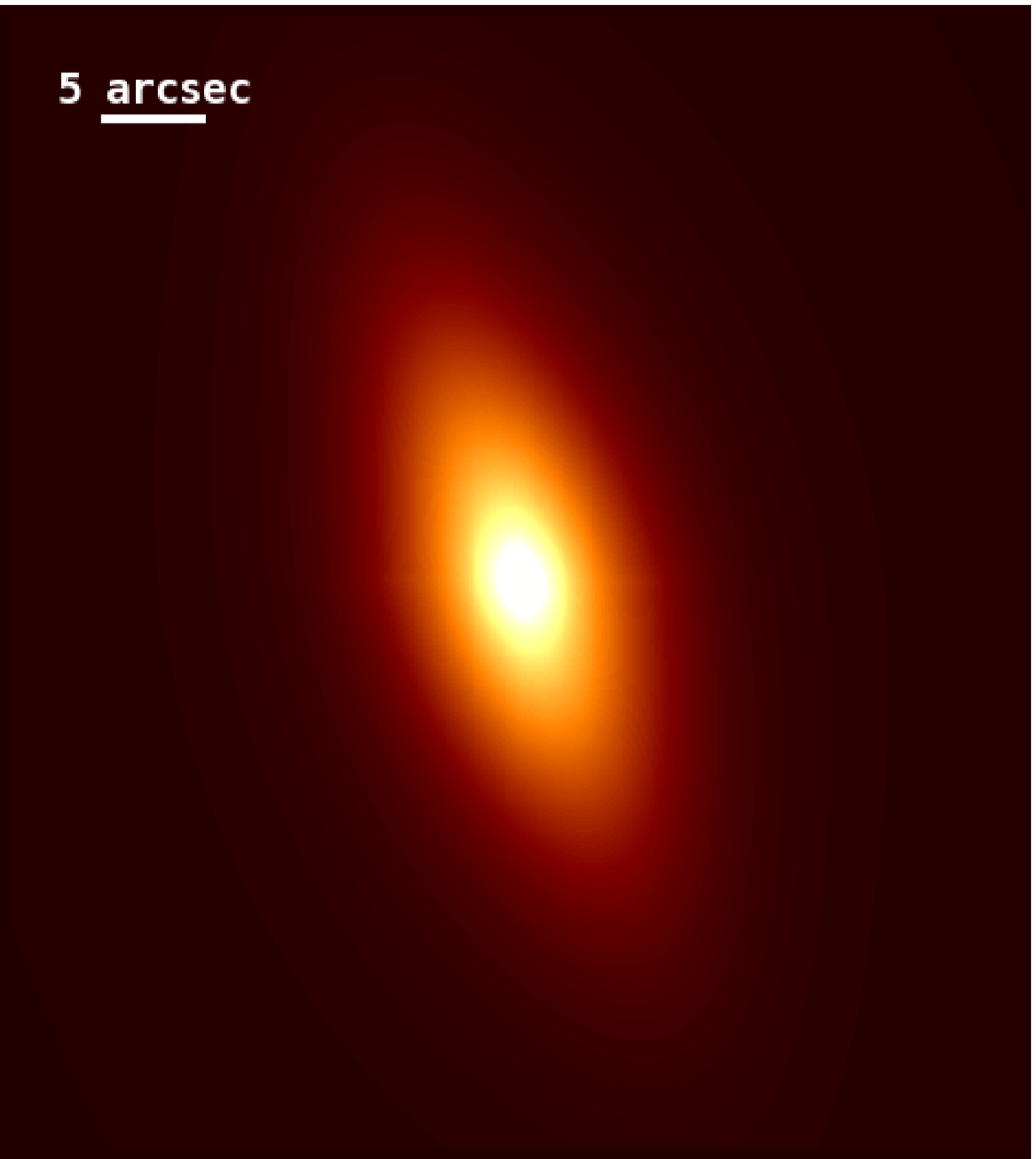} 
\includegraphics[trim=0cm 0cm 0cm -1.3cm, width=0.23\textwidth]{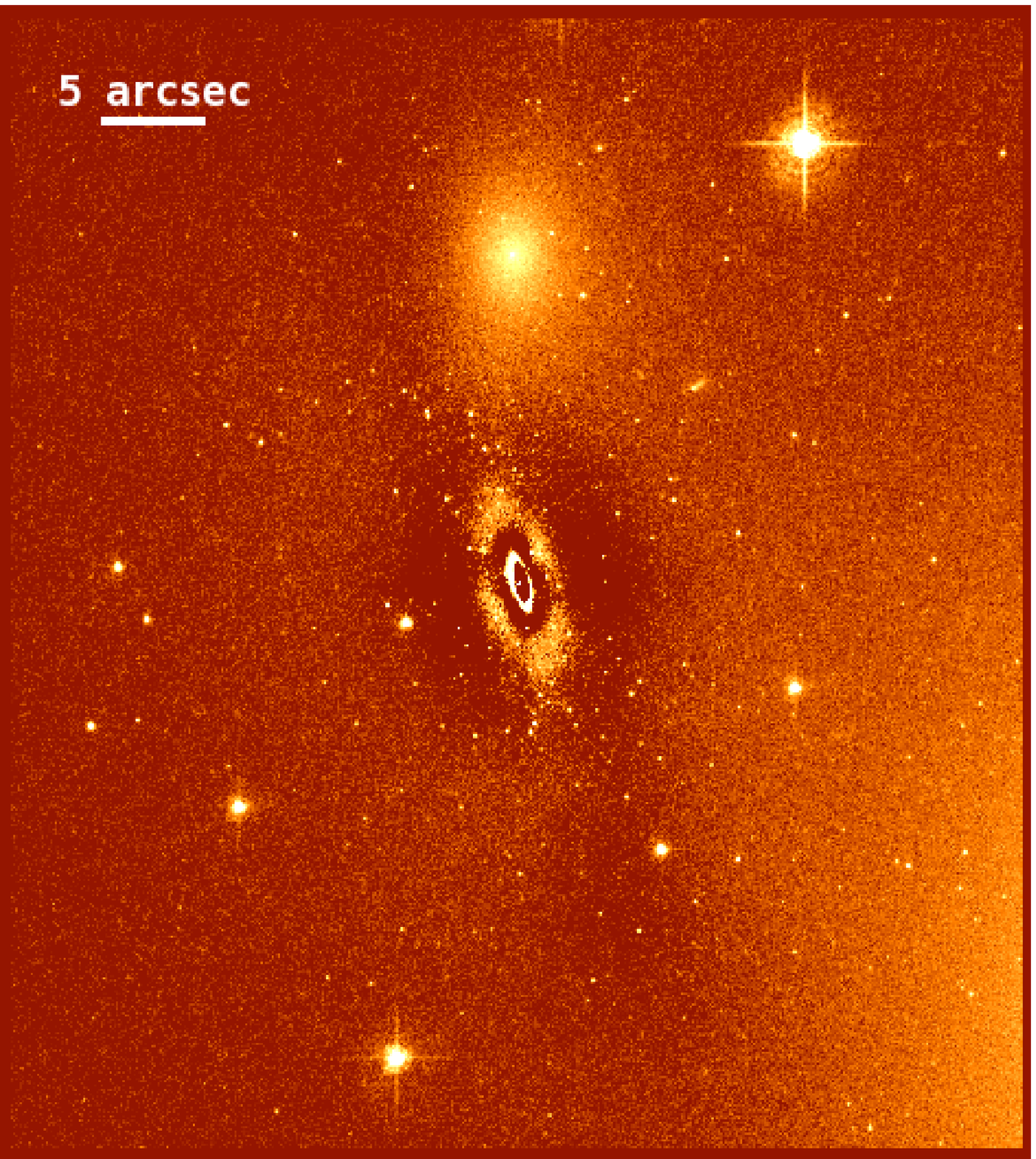} 
\caption{2D Model (left) and residuals (right). Due to the non-symmetrical nature of
  the dust ring in NGC~1277 (Figure~\ref{fig:ring}), we prefer the solutions to the 1D light
  profile (Figure~\ref{fig:fit} and Table~1). The off-centred, innermost ring seen here has a
  minor- and major-axis radius of  $\sim 0\arcsec .4$ and $\sim 1\arcsec .3$.} 
\label{fig:2D}
\end{figure}

Due to the non-symmetrical appearance of the dust in the inclined inner disk,
and the slightly differing nature of the light profiles along the major- and
equivalent-axis, our 2D model is not our preferred fit.  Due to the
symmetrical components that we used in {\tt Imfit}, as is commonly employed in
other 2D codes, we could not provide a perfect solution along every radial
direction.  Our 2D model and residuals can be seen in Figure~\ref{fig:2D}.
This decomposition resulted in an observed (i.e.\ not extinction nor dimming
corrected) spheroid magnitude equal to 14.28 mag (60\% of the total light), an
intermediate-scale exponential disk with an axis-ratio of 0.35 and a 
magnitude equal to 14.93 mag (33\% of 
the total light), and an inner inclined disk with a Gaussian magnitude equal
to 16.61 mag (7\% of the total light), amounting to a total observed magnitude
of 13.73 mag in the {\it F550M} (narrow $V$-band) filter, or 13.28 mag once
Galactic extinction corrected.
The axis ratio of the intermediate-scale disk corresponds to an inclination of
70$^{\circ}$, in good agreement with the result obtained from the geometry of
the dust ring reported at the end of the previous subsection.  This disfavours
the idea of an end-on bar in a more inclined disk.

\subsection{Color profile}\label{Sec_color}

\begin{figure}
\begin{center}
\includegraphics[width=1.0\columnwidth]{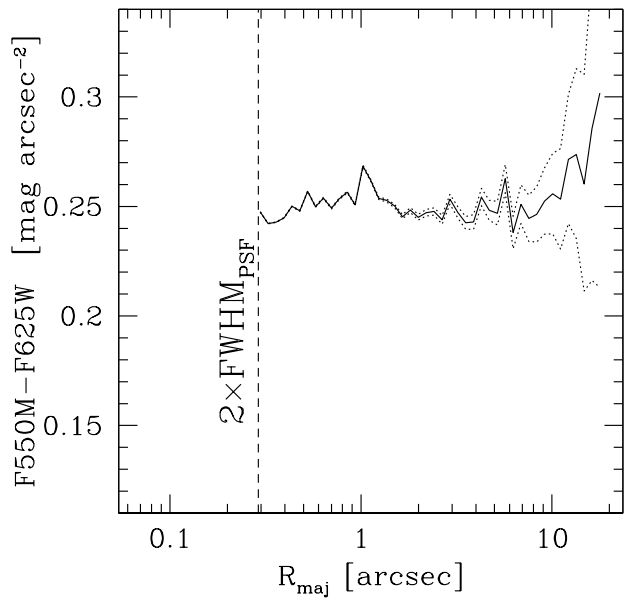} 
\caption{
Calibrated ({\it F550M}$-${\it F625W}) color profile for NGC~1277. 
}
\label{fig:col}
\end{center} 
\end{figure}

Due to the availability of an archived {\it F625W (SDSS r)} image, obtained
from the same {\it HST} Proposal Id.\ 10546 as the {\it F550M} image, it was
possible to construct a color profile for NGC~1277\footnote{We do not use the
{\it F625W} image as our primary image because the guide star
 acquisition failed on this exposure, which was also significantly shorter
 (1654 s) than planned and shorter than the 
{\it F550M} exposure (2439 s).}.  After checking the
alignment of this image with the {\it F550M} image, we ran \emph{IRAF}'s task
{\tt ellipse} on the {\it F625W} image in no-fit, photometry-only
mode\footnote{In this mode, the fitting algorithm is disabled and the task
  simply extracts photometry information from the image.}.  In this way, the
{\it F625W} surface brightness profile was extracted along the same ellipse
geometry as the {\it F550M} image.  Due to the different PSF in
these two images, we do not pay attention to the inner 2$\times$FWHM as given
by the {\it F625W} image.  Unfortunately, but not surprisingly, the closeness
of the two filters does not provide much useful information. 
The color profile shown in Fig.~\ref{fig:col} is 
largely consistent with no radial variation of the stellar population (at
least beyond the inner $\sim 1.\arcsec 5$).  The range of uncertainty shown in
Fig.~\ref{fig:col} by the dotted lines were derived by simultaneously adding
and subtracting the 1$\sigma$ uncertainty in the sky-background of each image.
The slight bump seen in the light profile from $\sim$0.4 to $\sim$1.5 arcseconds 
is likely due to the dust lane, although we do note that the spheroid light
contributes more than the intermediate-scale disk light over this radial
range. 

\begin{figure}
\begin{center}
\includegraphics[angle=90, width=1.0\columnwidth]{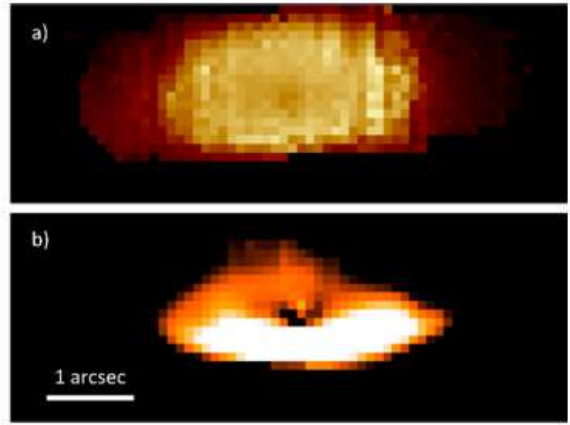} 
\caption{
Panel a): Flux at 2000 nm divided by the flux at 2350 nm, obtained
from the {\sc osiris} data cube.  The pixel size is $0\arcsec .1$ (35.2 pc). 
Panel b) Uncalibrated (2.2 $\mu$m $-$ {\it F550M}) color map emphasizing the inner dust
ring by setting the less reddened fluxes to black.  North is up and east is
to the left in both panels. 
}
\label{fig:ring}
\end{center} 
\end{figure}

Using a higher spatial-resolution, 2.2 micron image of just the inner few
arcseconds, obtained with the {\sc osiris} integral field unit on Keck I while using
the laser guide starm adaptive optics (see Section~\ref{Sec_Kin}), the {\it
  F550M}$-2.2 \mu$ colour reddens by more than 0.2 mag between $0\arcsec. 55$
and $1\arcsec .30$ along the major-axis, and peaks at 0.3 mag redder at 1$\arcsec$.
We have used this colour image to display the dust ring in Figure~\ref{fig:ring}. 

\section{Expected black hole mass} 
\label{Sec_predict} 

There are now a battery of scaling relations that can be used to
predict the masses of black holes at the centers of galaxies.  From
our decomposition of the galaxy light, and measurement of the
spheroid's stellar mass and central concentration of stars, we can
obtain two estimates of the expected black hole mass.  From the
velocity dispersion of the bulge, after avoiding biases from the
rotational kinematics of the inner disk, we can acquire a third
estimate.

\begin{table}[ht] 
\label{Tab_pred}
\centering
\caption{Expected black hole masses} 
\begin{tabular}{llc}
\hline\hline 
Scaling relation         &  Value                     &   Expected $M_{\rm bh}~(M_{\odot})$ \\
                          &                           &    [$10^9~M_{\odot}$]  \\
\hline
$M_{\rm bh} - n_{\rm maj}$   & $n=5.34$                   &  $0.57^{+1.29}_{-0.40}$ \\
$M_{\rm bh} - M_{\rm sph,*}$ & $M_{\rm sph,*}=2.69\times 10^{11}~M_{\odot}$  &  $1.58^{+4.04}_{-1.13}$ \\
$M_{\rm bh} - \sigma$      & $\sigma = 333$ km s$^{-1}$  & $4.40^{+8.19}_{-2.89}$ \\
$M_{\rm bh} - \sigma$      & $\sigma = 300$ km s$^{-1}$  & $2.27^{+4.04}_{-1.44}$ \\
\hline
\end{tabular}

Obtained using the S\'ersic $M_{\rm bh}$--$n_{\rm maj}$ relation from
  Savorgnan et al.\ (2013), the $M_{\rm bh}$--$M_{\rm sph,*}$ relation from
  Savorgnan et al.\ (2015) for early-type galaxies, and the $M_{\rm
    bh}$--$\sigma$ relation from Savorgnan \& Graham (2015a) for non-barred
  galaxies.  The uncertainties on
the black hole mass incorporate the uncertainties on the slope and zero-point of
the scaling relation, the intrinsic scatter in the relation, and the uncertainty in the
value used to derive the expected black hole mass.  See 
section~\ref{Sec_predict} for details and the use of older relations.
\end{table}

From our preferred 1D galaxy decomposition, and using the $M_{\rm
  bh}$--$n_{\rm maj}$ relation for ``S\'ersic bulges'' (i.e.\ those without
depleted cores) from Savorgnan 
et al.\ (2013, their Figure~4a and Table~4), the expected black hole
mass\footnote{See Graham \& Driver (2007) for the derivation of the
  uncertainty on the black hole mass, obtained here assuming an
  intrinsic scatter of 0.31 dex in the $M_{\rm bh} - n_{\rm maj}$
  relation.}, given a S\'ersic index of 5.34 and assuming a 20\%
uncertainty on $n_{\rm maj}$, is $\log M_{\rm bh} = 8.76 \pm 0.51$,
i.e.\ the optimal mass is $M_{\rm bh} = 5.74^{+12.88}_{-3.96} \times
10^8 M_\odot$.

Using the latest $M_{\rm bh}$--$M_{\rm sph,*}$ relation for bright early-type
galaxies from Savorgnan et al.\ (2015), which has a slope of 1.04$\pm$0.10,
and using $M_{\rm sph,*} = 2.69\times 10^{11}~M_{\odot}$, the expected black
hole mass\footnote{The uncertainty on the black hole mass given here assumes a
  spheroid mass uncertainty of 50\% and an intrinsic scatter of 0.42 dex in
  the $M_{\rm bh}$ direction of the $M_{\rm bh} - M_{\rm sph,*}$ relation.}
is such that $\log M_{\rm bh} = 9.20 \pm 0.55$, giving an optimal mass of
$M_{\rm bh} = 1.58^{+4.04}_{-1.13} \times 10^9 M_\odot$.  While the above
black hole mass is our preferred estimate when using the spheroid's stellar
mass --- due to the low level of scatter in the black hole direction about
this $M_{\rm bh}$--$M_{\rm sph,*}$ relation, and also due to the small
uncertainty on the slope --- for completeness we additionally report on the
expected black hole mass obtained when using the $M_{\rm bh}$--$M_{\rm
  sph,*}$ relation for S\'ersic galaxies.  The relation in Savorgnan et
al.\ (2015) gives $M_{\rm bh} = 2.86^{+12.30}_{-1.93} \times 10^9 M_\odot$.
Their S\'ersic $M_{\rm bh}$--$M_{\rm sph,*}$ relation has a slope of
1.48$\pm$0.20, which is smaller than the slope of 2.22$\pm$0.58 from the
relation in Scott et al.\ (2013) which predicts a black hole mass that is an
order of magnitude larger: $2.49 \times 10^{10} M_\odot$.  Perhaps
coincidentally, this value is in fair agreement with the black hole mass
of $1.7\pm0.3 \times 10^{10} M_\odot$ first reported by van den Bosch et
al.\ (2012).  However given the larger uncertainty on the slope of the
S\'ersic $M_{\rm bh}$--$M_{\rm sph,*}$ relation, this is not our preferred
estimate. 

Using the $M_{\rm bh}$--$\sigma$ relation for non-barred galaxies from
Graham \& Scott (2013, their Table 3) and assuming $\sigma = 333~\rm
km~s^{-1}$ (quoted from van den Bosch et al.\ 2012 outside the central
$1\arcsec .6$ -- their FWHM), the expected black hole mass\footnote{The
  uncertainties given here on the ($M_{\rm bh}$--$\sigma$)-derived
  black hole masses are based on an assumed 10\% uncertainty on the
  velocity dispersion, and 0.3 dex of intrinsic scatter in the
  $\log(M_{\rm bh}$ direction about the $M_{\rm bh}$--$\sigma$
  relation.} is $M_{\rm bh} = 2.78^{+3.98}_{-1.66} \times 10^9
M_\odot$.  Using the $M_{\rm bh}$--$\sigma$ relation for early-type galaxies 
from McConnell \& Ma (2013) gives $M_{\rm bh} = 3.48^{+5.64}_{-2.16} \times 10^9
M_\odot$, 
while using the $M_{\rm bh}$--$\sigma$ relation for non-barred
galaxies from Savorgnan \& Graham (2015a) gives $4.40^{+8.19}_{-2.89}
\times 10^9~M_{\odot}$
Repeating these estimations with $\sigma = 300$ km s$^{-1}$ 
(see the observed velocity dispersion map presented later) 
gives 
$M_{\rm bh} = 1.56^{+2.24}_{-0.93} \times 10^9 M_\odot$, 
$2.02^{+3.35}_{-1.24} \times 10^9~M_{\odot}$, and 
$2.27^{+4.04}_{-1.44} \times 10^9~M_{\odot}$, respectively.

For ease of reference, our preferred black hole mass estimates are presented 
in Table~2.

%

As detailed by Merritt \& Ferrarese (2001, see also Peebles 1972), 
the black hole's
sphere-of-influence can be regarded as the region of space where its gravity
dominates over that of other matter.  
Using $\sigma = 300$ km s$^{-1}$, a black hole mass of $2.27 \times 10^9 
M_\odot$ will have a sphere-of-influence radius 
$r_{\rm h} \equiv G M_{\rm bh} / \sigma^2 
           = 107.55 (M_{\rm bh}/10^9~M_{\odot})/(\sigma/ 200\, {\rm km\, s}^{-1})^2$ pc
= 109 pc $\simeq$0.31 arcsec, assuming a scale of 352 parsec per arcsecond. 
One obtains $r_{\rm h} = 0.49 \rm~arcsec$ when using $\sigma = 333$ km
s$^{-1}$ and $M_{\rm bh} = 4.40 \times 10^9 M_\odot$. 
These small values of $r_{\rm h}$ are consistent with Emsellem (2013) who
reported that there is no evidence of a black hole beyond 1.$\arcsec$6,
i.e.\ the FWHM of the PSF in the van den Bosch et al.\ (2012) kinematic data.

\section{Kinematic Data and Reduction}
\label{Sec_Kin}



The kinematic observations of NGC~1277 were taken with the OH Suppressing
InfraRed Imaging Spectrograph {\sc osiris} (Larkin et al.\ 2006) with the Keck
I Laser Guide Star Adaptive Optics (LGSAO) system (van Dam et al.\ 2006;
Wizinowich et al.\ 2006). {\sc osiris} is a lenslet array, integral field unit
spectrograph with a 2048$\times$2048 Hawaii-2 detector.  The spectral
resolution in the $K$-band (2.2 $\mu$m) with a 0.1 arcsec plate scale is about
3000, though this varies somewhat across the field.  We performed observations
of this object on the Hawaiian nights 23 and 24 November, 2013.  After a 60 second exposure
for target acquisition, each science data set consists of two 900-second
object frames, with an intervening 900-second sky frame in the standard
object-sky-object sequence. In all observations, the 0.100 arcsec per pixel
plate scale was used; at an `angular distance' of 72.5 Mpc, this gives a scale
of 35.2 pc per pixel (352 pc per arcsec).  We acquired 6 data sets on
November 23rd and 3 on November 24th, for a total of 4.5 hours on target, 
using the Kbb filter, which has a central wavelength
of 2180 nm and a bandwidth of 440 nm. The spectrometer was set at a position
angle of 90$^o$, roughly aligned with the long axis of the galaxy.  In all
observations, we used the laser guide star for the adaptive optics, using the
bright galaxy core as the tip-tilt guide star.

The observations were reduced as follows:
\begin{enumerate}
\item Each object frame was reduced using the standard {\sc osiris} data reduction
  pipeline (DRP)\footnote{{\sc osiris} Users Manual, Larkin et al., 2010, 
  \url{www2.keck.hawaii.edu/inst/osiris/OSIRIS\_Manual\_v2.3.pdf}} version~3.2, with the
  associated sky frames used for background extraction. The DRP uses
  ``Rectification Matrices'' which incorporate all the flat fielding, bias and
  dark frames in one, along with mapping each pixel to a lenslet/spectral
  element. These matrices are retrieved from the Keck repository; for our
  data, these were dated 6 June 2014.
\item After basic data reduction, the resulting data cube spectra still showed
  OH skylines, as the sky background was rapidly changing. The `Scaled Sky
  Subtract' DRP module (Davies 2007) was used to suppress these
  skylines; we measured dark frames on each night for this purpose.
\item The resulting data cubes have a field of view of 1.9$\times$6.4 arcsec
  (19$\times$64 pixels), with an extracted spectral range of 1965 nm to 2381
  nm at a spectral resolution of 0.25 nm. The values produced are in
  Analog-to-Digital Unit (ADU) 
  sec$^{-1}$.  Post data reduction, the FITS file imager and manipulating
  package QFITSVIEW\footnote{\copyright Thomas Ott 2011: \\
    \url{www.mpe.mpg.de/~ott/QFitsView/index.html}} with the underlying
  DPUSER language was used for analysis. The FITS files contain the primary
  data cube, with variance and data quality extensions. The FITS header WCS
  coordinate system was corrected to fix a known orientation issue. One of the
  data cubes had values and an image shape that was very dissimilar to the
  other 17 frames due to an unknown cause, so was discarded from the rest of
  the analysis. This problem was present in the original raw frame data, so it
  was not a problem generated by the DRP.
\item Telluric correction was applied using observations on the A0 star
  HIP16652 before and after each target visit, on both nights. These
  observations were also reduced by the DRP. The telluric spectrum is obtained
  by extracting the spectrum of the star; an aperture of about 5 pixels is
  used to acquire the total star light. Since hydrogen absorption lines
  dominate A0 star spectra, the strong Br$\gamma$ line was removed from the
  spectrum by fitting a Gaussian to the line profile. The resulting spectrum
  was divided by a blackbody curve at 9480 K, then normalized. The
  resulting spectrum showed strong broad absorption features in the 1990 to
  2080 nm range, with a secondary narrower absorption feature at 2317 nm. This
  spectrum is divided into the data cube of the galaxy at each spatial element
  to produce the corrected data cube.
\item The telluric corrected data cubes were mosaicked by finding the centroid
  of the 2D image created by summing along the spatial axis. On each night,
  the shift between successive frames in an observation block was usually less
  than one pixel; from the first to second night the shift was of the order of
  4 pixels. The cubes were shifted by integral pixel numbers in each spatial
  axis as required (sub-pixel shifts were not used, as they can produce
  interpolation issues). The final data cube is the averaged sum of the 17
  individual registered data cubes.
\item The DRP produces errant spikes and geometric data artifacts, which
  exhibit large data number transients (both positive and negative) for a
  single spatial and spectral element. These are identified as ``dead'' pixels
  and interpolated over using the QFITSVIEW functions dpixcreate and dpixapply
  on the spectrum at each spatial point. This cleaning was also applied to the
  telluric spectra before they were used.
\item We observed the following template stars (with spectral class noted),
  which were reduced and telluric corrected in the same manner as before, with
  the stellar spectrum extracted from the cleaned data cube; HD275038 (G5),
  HD275051 (K5), HD275052 (G5), HD275240 (M0), HD275246 (K5), HD275251 (G5) ,
  HD275337 (M0), HD275342 (K5) and HD275361 (M0). Each star was observed each
  night.
\end{enumerate}

\subsection{Measurement of Stellar Kinematics}

The stellar kinematics were extracted from the {\sc osiris} data cube using the
penalized Pixel Fitting ({\sc pPXF}) software\footnote{Available from: \\ 
\url{http://www-astro.physics.ox.ac.uk/~mxc/software/\#ppxf}} of 
Cappellari \& Emsellem (2004). This
technique constructs an optimal spectral template from a library of input
spectra, then convolves this optimal template with a line-of-sight velocity
distribution (LOSVD) to match the entire observed spectrum. This process is
iterated until the best-matching combination of weighted template spectra and
LOSVD is determined.

Although we observed template stars, the LOSVD map was found to be much more
regular, and with smaller formal uncertainties on all kinematic moments, when
using the Gemini spectral library of near-infrared late-type stellar templates
(Winge et al.\ 2009), which consists of 60 late-type stars observed with
either the GNIRS or NIFS instruments on the Gemini North
telescope\footnote{The template mismatch due to our small number of observed
  stars appears to be a much larger contribution to the uncertainty than any
  mismatch between the resolution or flux calibration of the library templates
  and the galaxy observation.}. This library is particularly suited to
extracting stellar kinematics from the CO band head features in the {\it
  K-}band.  All template spectra were smoothed to the {\sc osiris} spectral
resolution, $6.4 \AA$, as determined from off-target observations of narrow
night OH sky lines ($<0.1 \AA$ FWHM (Rousselot et al.\ 2000).

\begin{figure}
\begin{center}
\includegraphics[trim=0cm 7.1cm 0cm 6.5cm, width=1.\columnwidth]{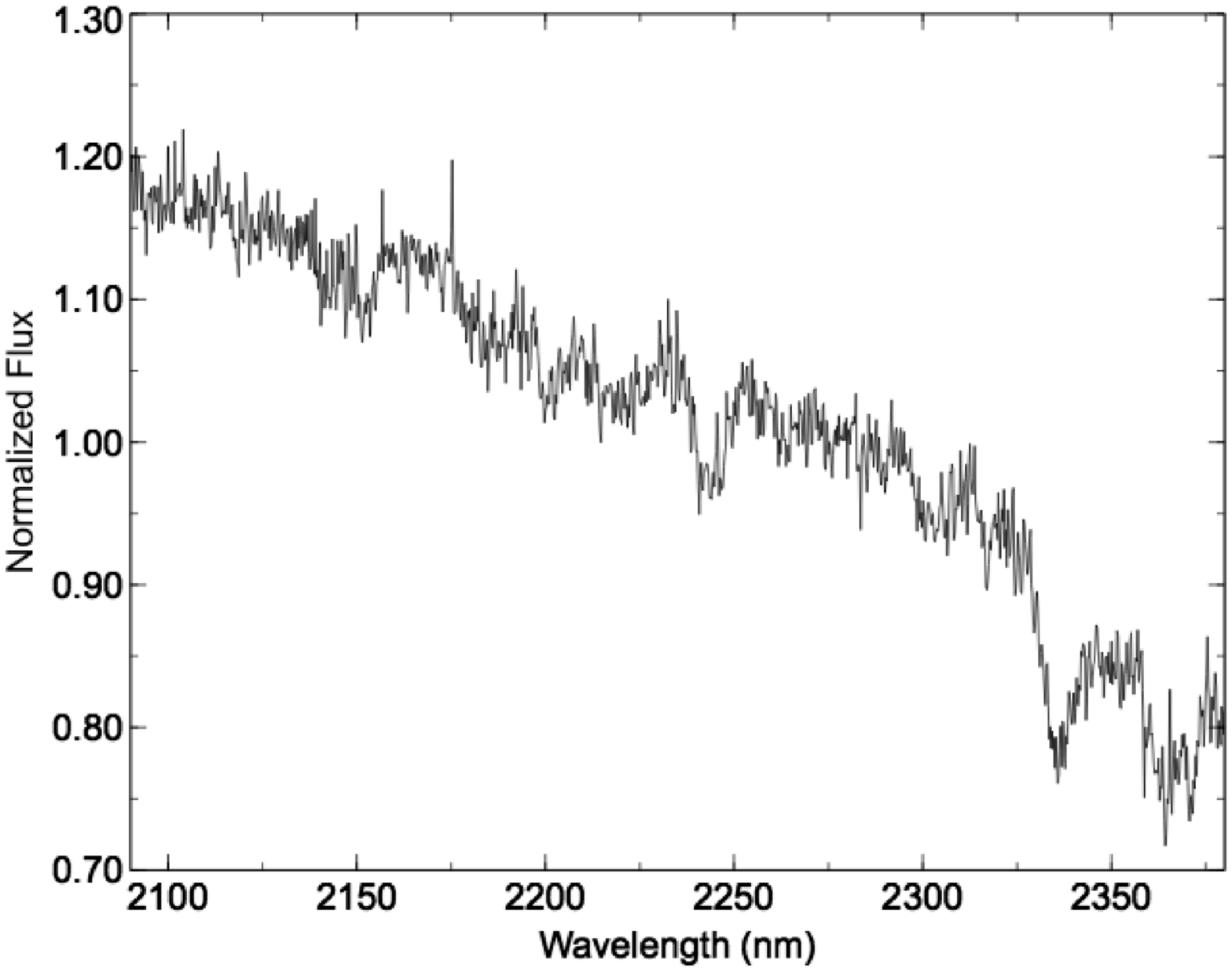}
\caption{High $S/N$ spectrum (black) of NGC~1277, extracted from the central
  1$\arcsec$ of the {\sc osiris} data cube.}
\label{fig:kinematic_fit}
\end{center}
\end{figure}

Initially we determined an optimal stellar template from a high
signal-to-noise ($S/N$) spectrum constructed by binning all spaxels within the
central 1$\arcsec$ of the {\sc osiris} data cube. We included a fourth-order
multiplicative polynomial in the fit to account for residual flux calibration
errors. We fit a LOSVD parameterized as a fourth-order Gauss-Hermite
polynomial, with the terms corresponding to the mean line-of-sight stellar
velocity, $v$, mean line-of-sight stellar velocity dispersion, $\sigma$, and
the third and fourth order Gauss-Hermite parameters $h_3$ and $h_4$,
respectively. The fit was penalized with a penalty $\lambda = 0.25$, following
Cappellari et al.\ (2011, their fig.~10), to appropriately bias the values of
$h_3$ and $h_4$ towards zero in spatial regions where the $S/N$ or $\sigma$ do
not allow us to constrain the higher-order shape of the LOSVD. In addition, we
made use of the {\sc CLEAN} algorithm (for details see Cappellari et al.,
2002), which iteratively removes pixels deviating more than 3$\sigma$ from the
template fit, repeating the process with the new set of good pixels until no
further bad pixels are identified. The number of pixels identified as bad in
this way is small relative to the length of the spectrum, and such pixels are
generally residual cosmic rays, emission, or artifacts on the CCD that were not
removed by earlier stages of the data reduction. This high $S/N$ central
spectrum is shown in Figure~\ref{fig:kinematic_fit}. 

We then determined the spatially resolved stellar kinematics of the central
regions of NGC~1277 by fixing this optimal template and convolving with a
LOSVD for each spaxel. We performed this both on individual spaxels (with $S/N$
ratio in the range $\sim$1--40) and on adaptively binned spaxels
(using the Voronoi technique of Cappellari \& Copin 2003), binned to a minimum $S/N$ of
40.  For both the binned and unbinned data, the resulting maps of stellar
velocity and velocity dispersion are entirely consistent, with the adaptively
binned data showing significantly less scatter in the outer parts of the
field-of-view as expected.

\begin{figure}
\includegraphics[width=0.235\textwidth]{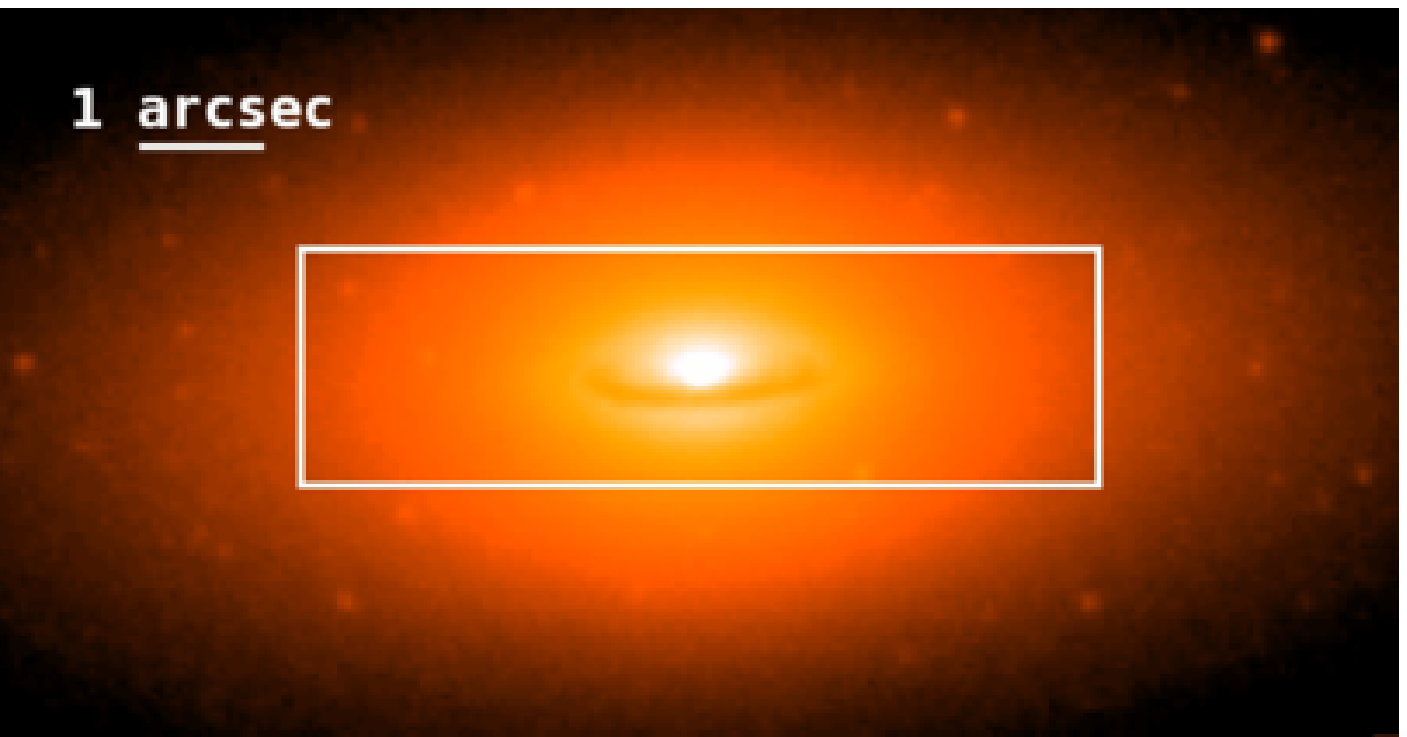}
\includegraphics[width=0.235\textwidth]{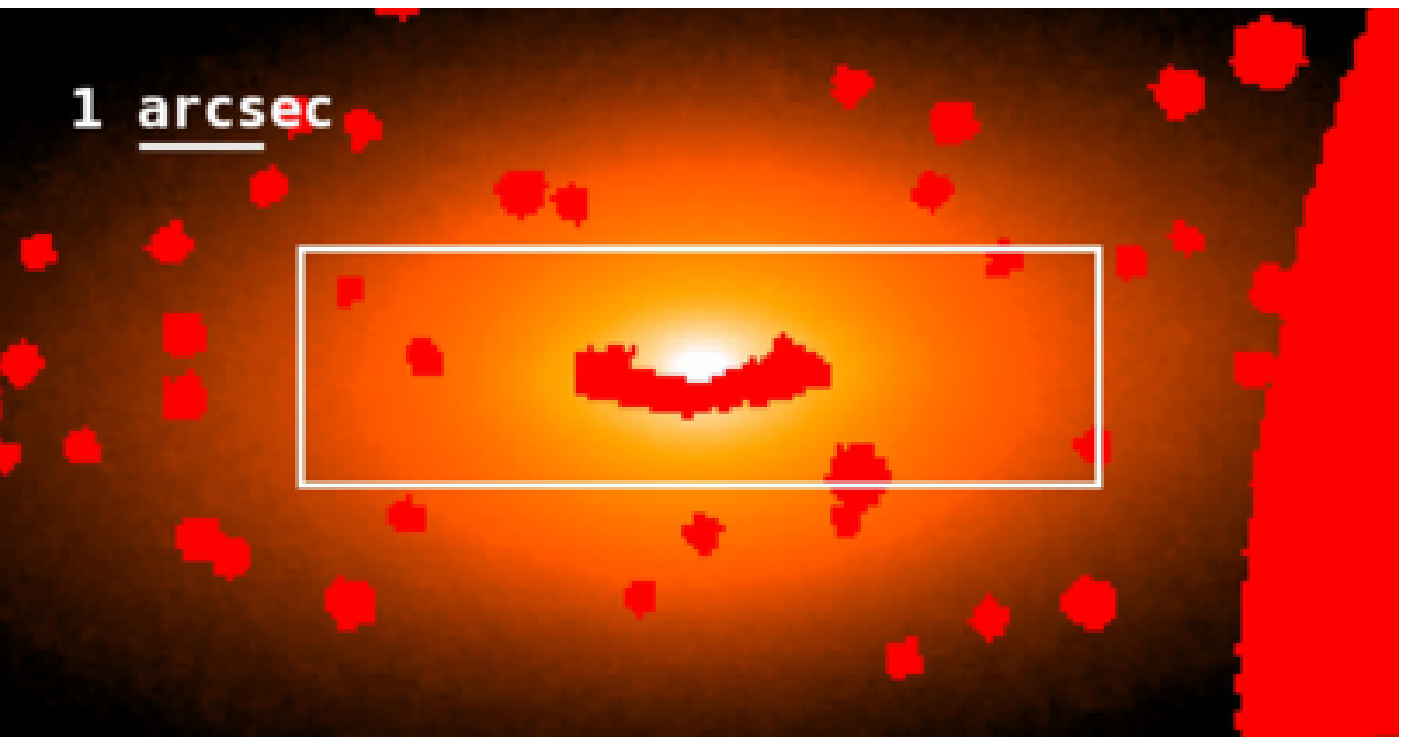} \\
\caption{The {\sc osiris} field-of-view is traced out by the rectangle
over an inner portion of the {\it HST} image (left panel) and the associated
masked image (right panel).  North is up, as in Figure~\ref{fig:ring}.}
\label{fig_trace}
\end{figure}

\begin{figure}
\begin{center}
\includegraphics[trim=0cm 2.8cm 0cm 0cm, width=\columnwidth]{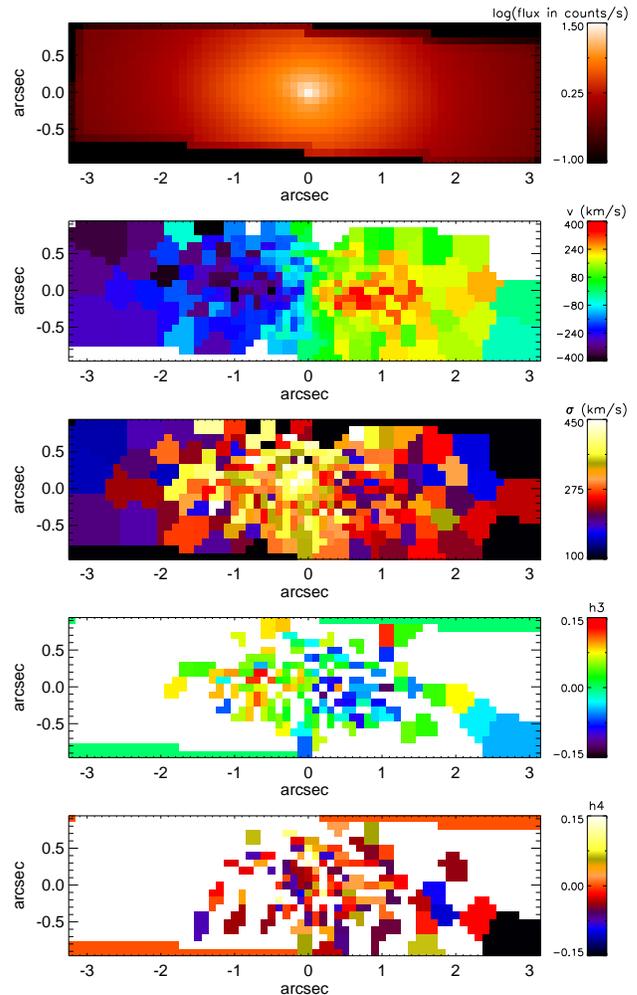}
\caption{Top panel: The (uncalibrated) 2.2 $\mu$m flux map from our {\sc
    osiris} data cube collapsed along the wavelength direction.  Second panel:
The (calibrated) velocity map after using Voronoi binning to a $S/N$ of 40.  
Third panel: The (calibrated) velocity dispersion map after using Voronoi
binning to a $S/N$ of 40.
Fourth and lower panel: The $h_3$ and $h_4$ maps, respectively, after masking
spaxels with $h_3$ and $h_4$ errors $>$ 0.05 in white. 
}  
\label{Fig:kin_maps}
\end{center}
\end{figure}

In Figure~\ref{fig_trace} we present the {\it HST F550M} image scaled to the
same size as the {\sc osiris} data, and show the details of the mask used to
exclude the effects of the inner dust disk.  In the upper panels of
Figure~\ref{Fig:kin_maps} we show the {\sc osiris} flux map, along with the
stellar radial velocity and velocity dispersion field in the central
$6.4\times1.8$ arcsec ($2250\times630$ pc) of NGC~1277.  The most striking
feature in this kinematic map is the steep velocity gradient seen along the
major-axis, which corresponds to the rotation of the nearly edge-on,
$1.8\times10^{10}~M_{\odot}$ inner disk seen over the same radial extent in
the photometric decomposition (Figure~\ref{fig:fit}).  The lower two panels in
Figure~\ref{Fig:kin_maps} show the $h_3$ and $h_4$ maps.  The anti-correlation
of $h_3$ with $v$ provides further evidence that there is an embedded disk in
NGC~1277, rather than a bar which would result in $h_3$ being correlated with
$v$ (see Bureau \& Athanassoula 2005). Furthermore, if there was a bar (other
than an end-on bar), the distribution of $h_4$ values would be negative rather
than scattered around zero as observed.  
The spaxels with larger errors in $h_3$ and $h_4$ 
generally coincide with regions of lower
$\sigma$, where the instrumental resolution has a greater impact on our
ability to determine the LOSVD. 
We note that the JAM model used in
the following section to determine the central black hole mass does not use
either the $h_3$ or $h_4$ parameters.

In the two neighboring spaxels on either side of the spaxel associated with the
kinematical centre of the galaxy (Figure~\ref{Fig:kin_maps}, second panel), 
the luminosity-weighted velocity changes 
from about $-200$ km s$^{-1}$ to about $+200$ km s$^{-1}$. There is thus a
large degree of rotational shear in the central spaxels, elevating their velocity
dispersion.  Indeed, in going from the mid-point of the above mentioned
spaxels, there is a velocity range of 400 km s$^{-1}$ across two
spaxels (0.2 arcseconds, 70 parsec).  With the 1$\arcsec$.6 (563 parsec) spatial
resolution available to van den Bosch et al.\ (2012), they were largely oblivious 
to this massive rotational motion boosting the velocity dispersion in their
central resolution element.  Around the inclined inner disk, the velocity
dispersion of the spheroid is 275 km s$^{-1}$ or less in many bins.
We have therefore used a rounded value of $\sim$300 km s$^{-1}$, 
in addition to the value of 333 km s$^{-1}$ (van den Bosch et al.\ 2012), 
but the true value may be lower.

\section{Measurement of the black hole mass} 
\label{Sec_Mass}

In this section we attempt to investigate if the new kinematic data 
might provide a black hole mass consistent with previous estimates. 
Following Emsellem (2013), we used a Multi-Gaussian photometric Expansions (MGE:
Monnet et al.\ 1992) and the Jeans Anisotropic MGE (JAM) modeling routine
(Emsellem et al.\ 1994; Cappellari 2008) to derive the black hole mass. 

For JAM modeling, one needs to estimate the spatial point-spread function (PSF)
of the kinematic data.  The PSF influences how much the black hole's kinematic
effect is smeared out and could have a significant effect on the resulting
measured mass.  In the simplest sense, PSFs from adaptive optics observations
can be split into two Gaussians: one has a width set by the natural
seeing disk while a second has a width set by the diffraction-limited 
core.  The quality of the Adaptive Optics (AO) correction sets the flux ratio
between these two Gaussians.  Using this framework we approximate our PSF as
the sum of two 2-dimensional Gaussians, which are also readily compatible with
MGE and JAM models.  Because our data consist of frames summed from two
different nights, we employ four Gaussian components: one pair of
seeing-limited and diffraction-limited for each night (normalized by exposure
time, 2:1)\footnote{With twice as much data obtained on the first night than
  was obtained on the second night, the lower $S/N$ of the second night's
  data (which had the better natural seeing) did not warrant using this data 
  alone.}.  The $K$-band seeing limits were estimated using the archival 
MASS/DIMM information from the Maunakea Weather Center
website\footnote{\url{http://mkwc.ifa.hawaii.edu/current/seeing/}}, which gave
an average full-width at half maximum (FWHM)\footnote{We took the average
  value from the Canada-France-Hawaii Telescope Weather Tower's MASS and DIMM
  seeing measurements at 0.5 $\mu$m (taken to be $1\arcsec .4$ and $0\arcsec
  .5$ on the first and second night) and multiplied by $(0.5/2.2)^{0.2}$ --- to
  adjust for the $\lambda^{-0.2}$ wavelength dependence --- to obtain the 2.2
  $\mu$m seeing.}  of 1.05 arcsec on 23 November 2013 and 0.37 arcsec on 24
November 2013.  In both cases, the diffraction-limit is set by the FWHM of the
inner Gaussian-like component of the Airy pattern, equal to 47
milli-arcseconds (derived from 1.03 $\lambda$/D, where $\lambda$ is the
observed wavelength 2.2 $\mu$m and D is the telescope diameter of 10
meters)\footnote{Although we used an {\sc osiris} plate scale of $0\arcsec .1$
  lenslets, our complete forward modelling convolves the model with the true
  PSF and then `under samples' the resulting model to compare with the
  observations.}.  The fraction of light in the diffraction-limited component
(measured by the `Strehl ratio') is allowed to vary.  Typical performance of
the Keck AO system suggests a Strehl ratio of $\sim$25\%.  Given that we do not
have a precise way to measure this for our data, we allowed for a wide 
range of Strehl ratios from 10\% to 40\%, but this had a fairly small 
affect ($\approx$15--20\%) on our black hole mass.

\begin{figure*}
\begin{center}
\includegraphics[width=\textwidth]{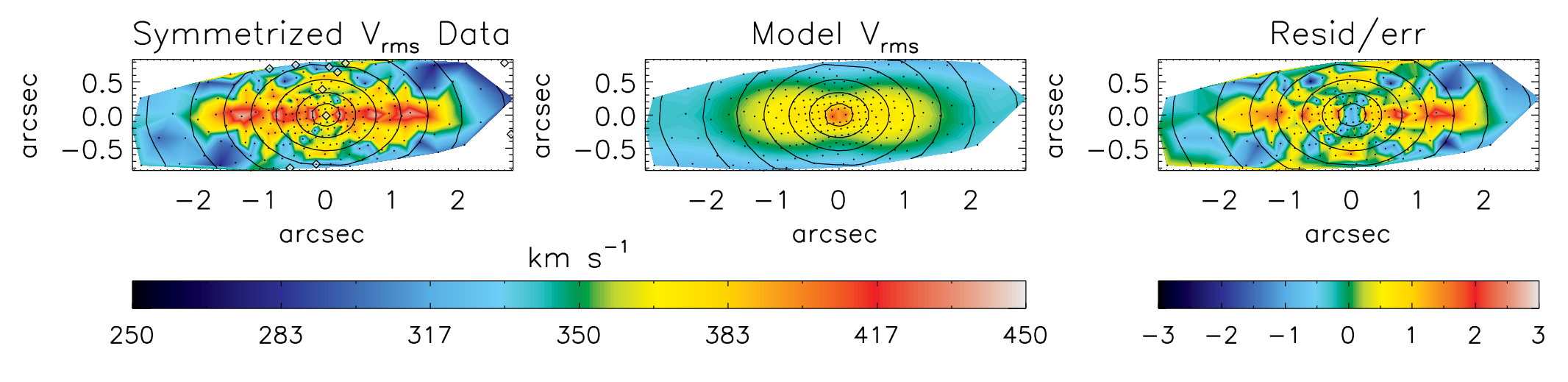}
\includegraphics[width=\textwidth]{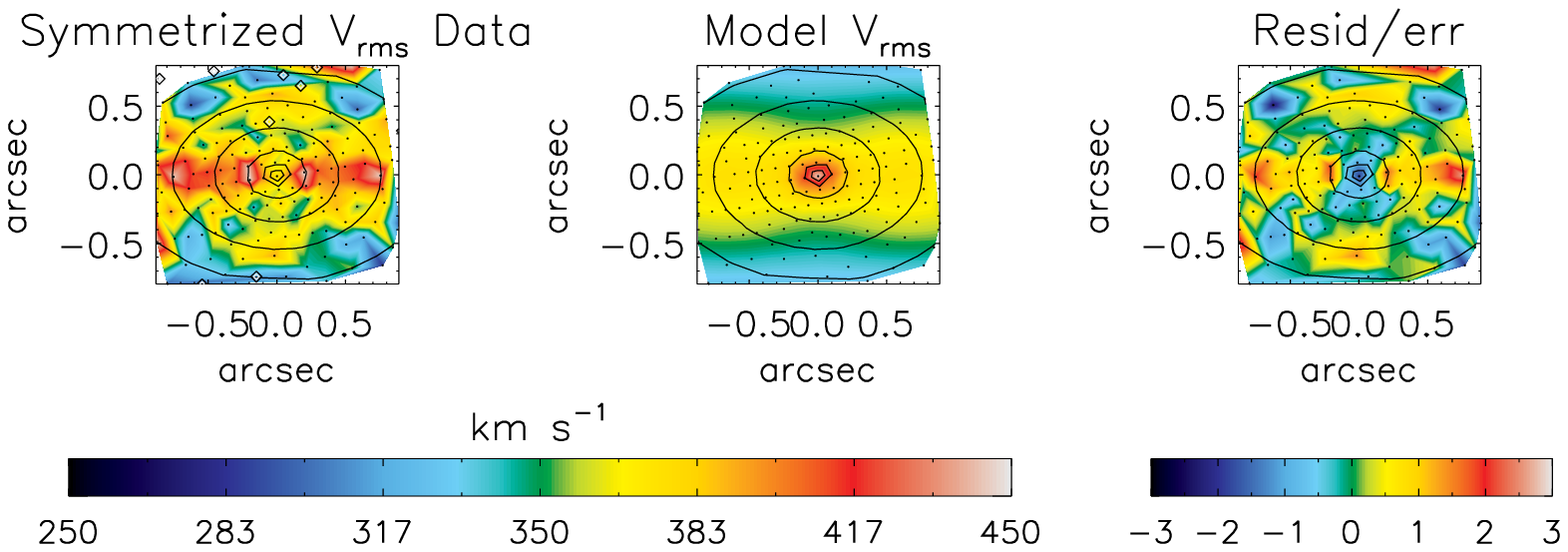}
\caption{
Top: 
JAM results for our MGE image reconstruction (see Table~\ref{Tab_MGE}),
showing the observed (symmetrized) $v_{\rm rms} = \sqrt{v^2 + \sigma^2}$ map
(left), the best fitting bi-symmetric $v_{\rm rms}$ model (center), and the
(data $-$ model) residuals scaled by the errors (right).  See
Section~\ref{Sec_Mass} for details.  Note: The data from
Figure~\ref{Fig:kin_maps} have been rotated by 8 degrees so that the inner
disk is now orientated horizontally.  If a 17 billion solar mass black hole
was present, then one would observe a symmetrical rise in $v_{\rm rms}$ within
the inner $\sim 1.\arcsec 6$.  Instead, we see that a rod-shaped feature
associated with the inner disk is responsible for the elevated $v_{\rm rms}$.
The JAM model's black hole removes the central portion of this, but this black
hole mass may be an over-estimate because it is fitting for both the elevated
dynamics around the black hole plus the central part of this rod-shaped
feature which the other components in the JAM model have not fully accounted
for.
Bottom: JAM results using kinematic data from 
just the inner $0.9\times0.9$ arcseconds. 
}
\label{JAMresults}
\end{center}
\end{figure*}

The measured black hole mass of a system obviously depends on how much mass is
attributed to stars in the central regions.  
As noted by Emsellem (2013), 
adequately masking out the region of the image which suffered too much extinction by the
central dust ring is important so that the MGE model does not under predict the
amount of light within the inner arcsecond. 
One should keep in mind that the dust in the disk does not simply obscure disk
light but also the light from the stars in the spheroid on the far side
of the dust disk.  This is substantial when the S\'ersic index is high and the
spheroid's stars are therefore considerably centrally concentrated. 
Comparing our photometrically uncalibrated {\sc osiris} 
$K$-band image (Figure~\ref{Fig:kin_maps}a) to the photometrically calibrated
optical {\it HST} image (Figure~\ref{fig_trace}) gives a relative estimate of
the dust extinction (see Section~\ref{Sec_color}).  We therefore 
began with the masked {\it HST/ACS F550M} image shown in
Figures~\ref{fig:image} and \ref{fig_trace}, and further masked out areas likely to
be affected by dust. 
%
%
We could not mask the {\it HST} image 
this heavily in subsection~\ref{Sec1D} because it caused the IRAF task {\tt
  ellipse} to fail. In that subsection we saw that the inner dust ring caused
the inner disk to be split into a central nucleus plus an inner ring in our
decomposition.  However here we are able to mask slightly more heavily.

We used the MGE model to match the (unmasked) light 
distribution seen in the {\it HST} image (see Table~\ref{Tab_MGE}), 
and allowed a spatially-constant 
mass-to-light ratio (M/L) to vary as a free parameter during the fitting 
process with the kinematic data. 
%
%
Y{\i}ld{\i}r{\i}m et al.\ (2015) report that there is evidence 
for negligible amounts of dark matter within the inner $\sim$3.6 kpc of
NGC~1277, and thus we have not included a dark halo component.

The best-fitting $M_{\rm bh}$, mass-to-light ratio, and orbital anisotropy
parameter $\beta$ are given in Table~\ref{Tab_fit}.  We ran Monte Carlo
simulations, refitting the JAM model 100 times to the best-fitting v$_{rms}$
map with different noise added according to the measurement errors.  The reported
uncertainties were calculated from the range of black hole masses that
encompassed 68\% of the resulting fits.  The close agreement between our
model's preferred $M/L_V$ ratio of 12.3 with the central ($r<1\arcsec .5$)
$M_*/L_V$ value of 11.65 from Mart\'in-Navarro et al.\ (2015a) is supportive of
this model's black hole mass of $1.2\times10^9~M_{\odot}$.  Fixing the Strehl
ratio to 25\% increased this mass to $1.4\times10^9~M_{\odot}$.  Using the
exact same MGE model from Emsellem (2013), based on his masking of the galaxy
image, coupled with our new kinematic data, gave $M_{\rm bh} = 1.8\substack{+0.3
  \\ -0.3}\times10^9~M_{\odot}$ and $M/L_V = 9.3$.  However using an MGE model built
with the incomplete dust mask from Figure~\ref{fig_trace} resulted in a black
hole mass of $12\times10^9~M_{\odot}$, dramatically emphasizing the point made by
Emsellem (2013) regarding the dust.
For reference, Emsellem (2013) reported an optimal black hole mass of
$5\times10^9~M_{\odot}$ and $M/L_V=10$. 
Our best-fitting 
JAM model gives a total stellar mass of $1.7\times10^{11} M_{\odot}$ based
on our input MGE model luminosity and the dynamical $M/L$ (=12.3) 
within the inner few arcseconds.  
This can be compared with the value of $1.8\times10^{11} M_{\odot}$ 
reported by Emsellem (2013) using $M/L=10$. 
The MGE models, consisting of Gaussian profiles (i.e.\ S\'ersic 
$R^{1/n}$ profiles with $n=0.5$) do not have the same extended envelopes as 
the high-$n$ light profiles observed out to large radii in massive early-type 
galaxies (Caon et al.\ 1993). Our MGE model flux is 43\% less than that of 
the component analysis 
which found a spheroid with a S\'ersic index of $\sim$5 and a total galaxy stellar
mass of $3.0\times10^{11} M_{\odot}$ (Table~1).  

\begin{figure*}
\includegraphics[angle=270, width=\textwidth]{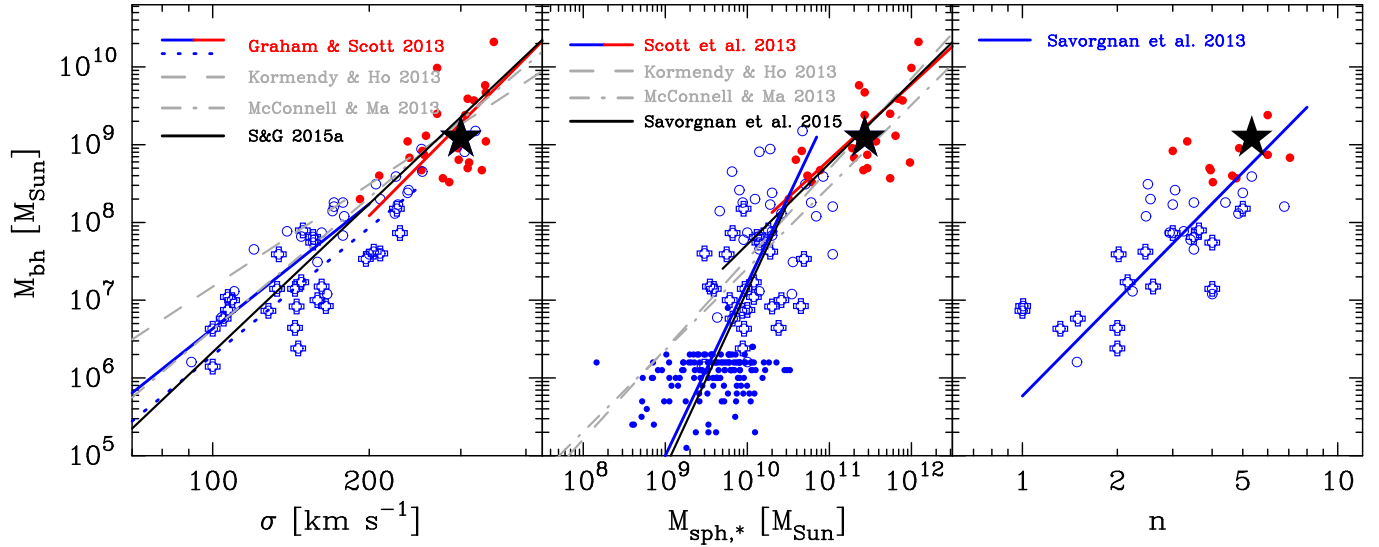}
\caption{
Left panel: $M_{\rm bh}$--$\sigma$ data and relations from Graham \& Scott
(2013): core-S\'ersic galaxies (red dots and red line);
unbarred S\'ersic galaxies (blue circles and blue line);
barred galaxies (blue crosses and blue dotted line).
Middle panel: $M_{\rm bh}$--$M_{\rm sph,*}$ data and relations from Scott
et al.\ (2013): core-S\'ersic galaxies (red line);
barred and unbarred S\'ersic galaxies (blue line).
The small blue dots denote AGN from Graham \& Scott (2015) but 
were not used to derive the blue line. 
Right panel: $M_{\rm bh}$--$n$ data and relation from Savorgnan et
al.\ (2013): barred and unbarred S\'ersic galaxies (blue line).
The black star denotes NGC~1277.  
Note: Kormendy \& Ho (2013) adjusted the central velocity dispersions to 
create reduced velocity dispersion estimates within $R_{\rm e}/2$ 
($\sigma_{R_{\rm e}/2}$) to which their shallower $M_{\rm bh}$--$\sigma$ 
relation pertains.  ``S\&G 2015a'' refers to the relation for unbarred
galaxies from Savorgnan \& Graham (2015a) while Savorgnan et al.\ (2015)
refers to their relation for late-type galaxies, derived using the modified
{\sc fitexy} routine from Tremaine et al.\ 2002 in a symmetrical manner. 
}
\label{Fig:compare}
\end{figure*}

Our best fitting $V_{\rm rms}$ model and residual map
is shown in the upper panels of Figure~\ref{JAMresults}. 
The striking rod-shaped feature --- which we 
fail to reproduce beyond the sphere-of-influence of the black hole --- is due to the 
rapidly rotating disk seen in the second panel of Figure~\ref{Fig:kin_maps}. 
Unfortunately our inability to model this inner disk casts doubt over the
reliability of the derived black hole mass plotted in
Figure~\ref{Fig:compare} for comparison with other galaxies. 
We varied the PSF such that it had (i) a Strehl ratio
of 100\%, i.e.\ a FWHM of $\sim 0\arcsec .05$ with no additional natural 
seeing component, and (ii) a natural seeing of $\sim 1\arcsec .2$ and a Strehl
ratio of 0\%.  We found that this had little effect on our ability to match
the rod-shaped feature (left panel of Figure~\ref{JAMresults}) and thus
remove it from the residual map (right panel of Figure~\ref{JAMresults}).
NGC~821 is another galaxy with a nearly edge-on
stellar disc for which the JAM modelling struggles to reproduce the strong
rod-like feature seen in the $V_{\rm rms}$ image (Cappellari 2008, his
Figure~5).  
This behaviour therefore appears to be symptomatic to the JAM model rather than
specific to NGC~1277. 
We then reduced the `kinematic field-of-view' used by the JAM
model by including only the central $0\arcsec .9\times0\arcsec .9$ of
kinematic data.  This was done to match the `kinematic field-of-view' used by
Walsh et al.\ (2015), but had no significant effect on our results, giving 
$M_{\rm bh}=1.2^{+0.4}_{-0.3}\times10^9~M_{\odot}$, $M_*/L_V = 12.4$,
$\beta=0.03$ and a Strehl ratio of 40\% (see Figure~\ref{JAMresults}). 

We conclude that the JAM model, as used by us, is challenged by the edge-on
nature of the inner disk in NGC~1277, {\it perhaps} in a scenario analogous to
the isophotal fitting problems faced by the IRAF routine {\tt ELLIPSE}
(Jedrzejewski 1987) which has recently been remedied using a Fourier series to
describe perturbations from pure elliptical isophotes (Ciambur 2015).  The
non-elliptical contours of the image and/or the velocity dispersion map may
partly be the source of the residual disk seen in Figure~\ref{JAMresults}, but
resolving this is beyond the scope of the present paper.  We note that recent
testing of the JAM model (Li et al.\ 2015) has not yet explored this issue.
It may more simply be that we need to construct models with more massive, thin
planar disks that {\it can} explain the high v$_{rms}$ extending out to
$\sim$2$\arcsec$.  Due to this, we report a {\it tentative} black hole mass
that we hope to refine in future work by better accounting for the elevated
v$_{rms}$ beyond the sphere-of-influence of the black hole, due to the near
edge-on disk.  The bias that this introduces to our current work is in the
sense that we may have over-estimated the black hole mass to account for some
of the increased v$_{rms}$ which is due to the inner disk.

\section{Discussion and Conclusions}\label{Sec_DC}

We have used the kinematics, the ellipticity profile, the light profile, and
the `fourth harmonic' profile describing the deviation of the isophote contours
from perfect ellipses, to obtain a physically-sound and consistent description
of the structural components in NGC~1277.  
This was called for in Emsellem (2013).  Contrary to the claim in van den
Bosch et al.\ (2012) that ``a classical bulge appears to be entirely absent''
in NGC~1277 (because they obtained S\'ersic indices less than 2 from their
fitting of four galaxy components) --- a claim they re-iterate in
Y{\i}ld{\i}r{\i}m et al.\ (2015) --- we find that NGC~1277 has a massive bulge
with a major-axis S\'ersic index of 5.34.  This dominant spheroidal component
accounts for 79\% of the light and 90\% of the mass in NGC~1277, weighing in
at $M_{\rm sph,*} = 2.69\times10^{11}~M_{\odot}$.
This increase in the spheroidal component's stellar mass, 
above the value of 0.29$\times10^{11}~M_{\odot}$ from van den
Bosch et al. (2012), is mostly due to the galaxy decomposition but is also because
they used a $V$-band mass-to-light ratio of 6 while we used 11.65. 
We find that the spheroidal component 
has a half light radius of 6$\arcsec$, rather than $\sim$1$\arcsec$ (van den
Bosch et al.\ 2012), which may help to resolve the puzzle in 
Mart\'in-Navarro et al.\ (2015a) as to why, at 1.5$\arcsec$ (previously thought to be 1.5
$R_{\rm e,sph}$) the stellar `initial mass function' (IMF) had not declined and
  was still bottom heavy at 1.5$\arcsec$, in contrast to results beyond
  1$R_{\rm e,sph}$ in other high-mass 
  galaxies (Mart\'in-Navarro et al.\ 2015b; La Barbera et al.\ 2015). 
Of course, if the IMF of the spheroidal component
of NGC~1277 is found to change at large radii, 
then the stellar mass-to-light ratio and the total stellar mass of
the spheroid will change. 

\begin{table} 
\label{Tab_MGE} 
\centering
  \caption{Multi-Gaussian Expansions (MGE) model parameters.}
\begin{tabular}{lrc}
\hline
\hline
  $\Sigma_{0}$            &  $\sigma_{G}$   &  q  \\
 $L_{\odot}$ pc$^{-2}$    &   [arcsec]      &     \\
\hline
      160183.  &   0.0472 &     0.666 \\
      17367.3   &  0.2391 &     0.745  \\
      7196.26   &  0.6507 &     0.586  \\
      3014.67   &  1.3140 &     0.677  \\
      797.645   &  3.9523 &     0.392  \\
      129.483   &  11.0372 &    0.414  \\ 
\hline
\end{tabular}

Column 1: Maximum central surface brightness for each two-dimensional (projected)
Gaussian.  Column (2) provides the associated width, and column (3) 
gives the axis ratio. 
\end{table}

\begin{table} \label{Tab_fit}
\centering
  \caption{Jeans Anisotropic MGE (JAM) result}
\begin{tabular}{cccc}
\hline
\hline
 $M_{\rm bh}$                 &  $M/L_V$                   &  $\beta$  &  Strehl \\
 $10^9~M_{\odot}$             & $M_{\odot}/L_{\odot, V}$ &           &  ratio  \\
\hline
 $1.2\substack{+0.3 \\ -0.3}$ &     12.3                   &  0.03     &   40\%   \\
\hline
~
\end{tabular}
\end{table}

The abstract of van den Bosch et al.\ (2012) refers to this galaxy as a
compact lenticular galaxy with a total stellar mass in excess of
$10^{11}~M_{\odot}$.  Trujillo et al.\ (2014) has similarly identified this
galaxy as a likely descendant of the compact massive galaxies seen at $z \sim
2\pm0.5$, and several other likely descendants were previously identified in
Dullo \& Graham (2013).  NGC~1277 may indeed have been a compact massive
galaxy at $z \sim 2$ which had, or has since acquired, an intermediate-sized
stellar disk.  The disk accretion scenario could explain the decreased stellar
mass-to-light ratio as one transitions from the spheroid dominated inner
region of the galaxy to the disk-dominated region at intermediate radii
(Mart\'in-Navarro et al.\ 2015a).  One might therefore wonder that if the
spheroid in NGC~1277 has not evolved, then perhaps the (black
hole)-to-spheroid mass ratio has also not evolved.  One may of course also
wonder if NGC~1277 previously boasted a larger disk, which has since been
eroded away in its Perseus cluster environment.

The observed mass-scaling between $M_{\rm bh}$ and the many properties of the
host galaxy/spheroid (Hutchings et al.\ 1984; Dressler 1989; Yee 1992;
Kormendy \& Richstone 1995; Laor 1998; Magorrian et al.\ 1998; etc., see
Graham 2015 for a historical review) 
  %
  %
is well known to suggest a fundamental connection between massive black holes and their
host spheroids.  Such observations also provide key input into theoretical
investigations trying to determine the details of black hole and galaxy
evolution (e.g., Ciotti \& van Albada 2001; Adams et al.\ 2003; Heckman et
al.\ 2004; Cattaneo et al.\ 2005; Cirasuolo et a.\ 2005; 
Fontanot et al.\ 2006, 2015; Robertson et
al.\ 2006; Shankar et al.\ 2006, 2009; 
Treu et al.\ 2007; Ciotti 2008).  
For reference, Savorgnan et
al.\ (2015) reveals that the $M_{\rm bh}/M_{\rm sph,*}$ mass ratios range from
0.1--5\% in (non-dwarf) early-type galaxies. 
This places an important emphasis
on any (black hole)-(host spheroid) system that appears to not conform with
these relations, as the evolution of those systems may need a separate 
explanation (Ferr\'e-Mateu et al.\ 2015). 



Our AO-assisted Keck I {\sc osiris} data suggests the presence of a
$1.2\pm0.3\times10^9~M_{\odot}$ black hole in NGC~1277 and
$M/L_{V,sph}=12.3~M_{\odot}/L_{\odot}$, where these errors only consider
random noise in the kinematic data and do not account for any systematic
errors.  This black hole mass may be an upper limit given the negative 
residual at the centre 
of Figure~\ref{JAMresults}, however we need to further refine our analysis to
better account for the rod-shaped feature seen beyond the black hole's
sphere-of-influence ($0\arcsec .2$) in this $V_{\rm rms}$ residual map.  Walsh
et al.\ (2015) report a black hole mass of $4.9\pm1.6\times10^9~M_{\odot}$ and
$M/L_{V,sph}=9.3\pm1.6~M_{\odot}/L_{\odot}$ (in agreement with the optimal
solution in Emsellem 2013).  From Figure~2 in Walsh et al., one can see that
if $M/L_{V,sph}=11.65~M_{\odot}/L_{\odot}$ then one would have $M_{\rm bh}
\approx 4\times10^9~M_{\odot}$, and at $M/L_{V,sph}=12.3~M_{\odot}/L_{\odot}$
one has $M_{\rm bh} \approx 3\times10^9~M_{\odot}$. That is, the degeneracy
between the black hole mass and the stellar mass-to-light ratio can {\it
  partly} explain some of the differences between us and Walsh et al.\ (2015).
Of course other differences exist, such as our problematic JAM modeling of the
inner $\sim 3\arcsec\times1\arcsec$ (radius) containing the edge-on disk
versus their Schwarschild modeling of just the inner $\sim
1\arcsec\times1\arcsec$ (radius).  Coupled with the host spheroid mass, these 
two black hole masses (i.e.\ ours, and that from Emsellem et al.\ 2012 and
Walsh et al.\ 2015) give a (black hole)-to-(host spheroid) mass ratio of
0.45 and 1.8 percent.  We therefore conclude that NGC~1277 does not host an
over-massive black hole.  This downward revision of this ratio for NGC~1277 may
not bode well for recent papers claiming NGC~1277 as a potential prototype in
support of the direct collapse of massive black holes (e.g.\ Bonoli et
al.\ 2014; Tanaka \& Li 2014).


%

NGC~4342 is another lenticular galaxy without a partially-depleted core and 
with an allegedly over-massive black hole (Cretton \& van den Bosch 1999) --- 
weighing in at nearly 7 per cent of the bulge mass (Bogd\'an et al.\ 2012b).
While Blom et al.\ (2014) reveals that some of this galaxy has been stripped
away by its neighbour, which would raise the $M_{\rm bh}/M_{sph}$ mass ratio,
and Valluri et al.\ (2004) highlight concerns with its estimated black hole mass, 
of additional relevance is the presence of an inner stellar disk in this
galaxy (van den Bosch, Jaffe, \& van der Marel 1998).  This galaxy has a
strong central 
rotation reaching 200 km s$^{-1}$ by r=0.25 arcseconds.  Furthermore, like
NGC~1277, this galaxy (along with NGC~4570) was noted by van den Bosch et
al.\ (1998) to be somewhat intermediate between an elliptical and a lenticular
galaxy.  
In addition, NGC~4342 has an old age of around 8 Gyr (van den Bosch et 
al.\ 1998), and both NGC~4342 and NGC~1277 additionally display minimal
central X-ray emission (Bogd\'an et al.\ 2012a; Fabian et al.\ 2013).  
Given this large number of similarities, 
see also Emsellem et al.\ (2013) in regard to the kinematic similarity, 
we speculate that rotational velocity shear 
(primarily due to the self-gravity of the inner disk rather than a black
hole) or non-elliptical contours arising from the nearly edge-on
disk, may have increased the derived black hole mass in NGC~4342. 
Indeed, Valluri et al.\ (2004) highlighted that past data supporting such a
large, and indeed any, black hole in NGC~4342 was questionable.  

There are yet other galaxies with inner disks that may also be worth further
investigation. For example, 
the nuclear disk (plus central star cluster) in NGC~7457 
(Balcells et al.\ 2007) similarly casts doubts over this galaxy's black 
hole mass (Gebhardt et al.\ 2003; Schulze \& Gebhardt 2011). 
The nuclear disk in NGC~4486B (Lauer et 
al.\ 1996; Gu\'erou et al.\ 2015, see their Figure~2) 
may have similarly contributed to the reportedly high $M_{\rm bh}/M_{\rm sph,*}$ mass 
ratio of 9\% (Kormendy et al.\ 1997) in this galaxy. 
Higher spatial resolution spectroscopic data than {\it HST} can provide may be
required to resolve this issue in these relatively small, low-luminosity
galaxies.  The upcoming, extremely large ground-based telescopes will be key
players in this regard (e.g.\ Do et al.\ 2014).  Refinements to the 
 Jeans Anisotropic MGE (JAM) modeling routine is also expected to be helpful
 and will be presented in a forth-coming paper. 
Nonetheless, we report that the spheroid mass in NGC~1277 is nearly an order
of magnitude greater than previously realized, and its black hole mass appears
to be an order of magnitude smaller after better resolving the rapidly
rotating, edge-on inner disk.  This dramatically reduces the $M_{\rm
  bh}/M_{\rm sph,*}$ mass ratio in NGC~1277. 

\acknowledgments

We wish to thank Eric Emsellem for his input to this work.  
This research was supported under the Australian Research Council’s
funding scheme (DP110103509 and FT110100263), and through 
Swinburne's Keck time Project Code W161OL.
This work is based on observations made with the NASA/ESA Hubble Space
Telescope. 
Further support was provided by proposal number HST-GO-11606 (PI: Batcheldor) 
awarded by NASA through a grant from the Space Telescope Science Institute, which is
operated by the Association of Universities for Research in Astronomy,
Incorporated, under NASA contract NAS5-26555.  
Some of the data presented herein were obtained at the W.M.\ Keck
Observatory, which is operated as a scientific partnership among the
California Institute of Technology, the University of California and the
National Aeronautics and Space Administration. The Observatory was made
possible by the generous financial support of the W.M.\ Keck Foundation.
The authors wish to recognize and acknowledge the very significant cultural
role and reverence that the summit of Maunakea has always had within the
indigenous Hawaiian community.  We are most fortunate to have the opportunity
to conduct observations from this mountain.

\end{document}